\documentclass[%
 preprint,
 superscriptaddress,
 amsmath,amssymb,
]{revtex4-2}

\usepackage{graphicx}
\usepackage{dcolumn}
\usepackage{bm}
\usepackage{hyperref}
\usepackage{cleveref}
\usepackage{siunitx}
\usepackage{mathtools}
\usepackage{longtable}
\usepackage{amsfonts}

\def\doubleunderline#1{\underline{\underline{#1}}} 

\begin{document}

\title{Exploring the stability of twisted van der Waals heterostructures}

\author{Andrea Silva}
\email{a.silva@soton.ac.uk}
\affiliation{Engineering Materials, University of Southampton}
\affiliation{national Centre for Advanced Tribology Study at University of Southampton}

\author{Victor E.P. Claerbout}
\email{claervic@fel.cvut.cz}
\affiliation{Advanced Materials Group, Department of Control Engineering, Faculty of Electrical Engineering, Czech Technical University in Prague (CTU)}

\author{Tomas Polcar}
\affiliation{Engineering Materials, University of Southampton}
\affiliation{national Centre for Advanced Tribology Study at University of Southampton}
\affiliation{Advanced Materials Group, Department of Control Engineering, Faculty of Electrical Engineering, Czech Technical University in Prague (CTU)}

\author{Denis Kramer}
\affiliation{Engineering Materials, University of Southampton}
\affiliation{Mechanical Engineering, Helmut Schmidt University, Hamburg, Germany}

\author{Paolo Nicolini}
\affiliation{Advanced Materials Group, Department of Control Engineering, Faculty of Electrical Engineering, Czech Technical University in Prague (CTU)}

\begin{abstract}
  Recent research showed that the rotational degree of freedom in stacking 2D materials yields great changes in the electronic properties. Here we focus on an often overlooked question: are twisted geometries stable and what defines their rotational energy landscape? Our simulations show how epitaxy theory breaks down in these systems and we explain the observed behaviour in terms of an interplay between flexural phonons and the interlayer coupling, governed by Moir\'e superlattice. Our argument applied to the well-studied MoS$_2$/Graphene system rationalize experimental results and could serve as guidance to design twistronics devices.
\end{abstract}

\maketitle

\section*{Introduction}
Van der Waals (vdW) 2D materials, like graphite and the family of transition metal dichalcogenides (TMDs), are a class of compounds characterized by an anisotropic structure.
Strong intralayer covalent bonds complement the weak vdW interlayer interactions which facilitate the lamellar structure of bulk crystals.
Due to their diverse chemistry and versatile properties, these materials have received significant attention from the scientific community in the past decades \cite{mannix2017,Chhowalla2013,Vazirisereshk2019}.
Applications can be found in microelectronics \cite{radisavljevic2011,lopezsanchez2013}, catalysis \cite{Lauritsen2004,shi2009a,shi2009} and tribology \cite{dienwiebel2004, Vazirisereshk2019a}.
While the attractive properties of the pure compounds are widely known, recent efforts have been focusing on the physics and properties emerging from the stacking degree of freedom offered by these lamellar materials.
Different types of single layers can be mixed and matched to create new superstructures, termed \textit{heterostructures}~\cite{geim2013,novoselov2016a,liu2016,li2019}.
A key feature affecting the behaviour of multi-layered structures is the relative orientational mismatch between layers.
While heterostructures are intrinsically incommensurate due to the different lattice constants of the parent single layers,  incommensurability can also arise in homostructrures due to relative misalignment of the single layers~\cite{dienwiebel2004}.

The relative mismatch between layers, both for homo- and heterostructures, has been related to a range of electronic and mechanical properties~\cite{zhu2019,du2017,lihuang2015,dienwiebel2004,martin2018}.
A flourishing new branch in the field of condensed matter, known as \textit{twistronics}, promises to allow fine-tuning of the electronic properties using the rotational misalignment between layers \cite{lihuang2015,cao2018}.
A notable example is the recent discovery of unconventional superconductivity in bilayer graphene (G) twisted at the \textit{magic angle} of \ang{1.1} \cite{cao2018}.
Another study found that the vertical conductivity of bilayer MoS$_2$/G heterostructures varies by a factor of five when imposing an angle of \ang{30} between the layers \cite{liao2018}.
Finally, a pioneering work~\cite{dienwiebel2004} showed that, by switching from commensurate to incommensurate orientation in graphite systems, it is possible to achieve a state in which the coefficient of friction vanishes, the so-called \textit{superlubric} regime.

Despite the interesting physics that results from these relative mismatches, an often overlooked question is what determines their rotational energy landscape and thus which geometries are stable.
Indeed, experimental studies are contradictory on this point, with a wide range of misfit angles found, even for the same type of system \cite{Liu2016a,shi2012,lu2015,adrian2016}.
Below we give a few examples of heterostructures based on MoS\textsubscript{2} on G.
This system may be regarded as the prototypical 2D heterostructure, as it combines two well-known and extensively studied materials, widely reported on in the literature.
Using chemical vapor deposition (CVD), Liu~\textit{et~al.} epitaxially grew triangles of MoS\textsubscript{2} on top of G, about $\SI{0.135}{\mu m}$ in size, with the majority of them (84\%) aligned to the substrate and the remainder rotated by $\ang{30}$ \cite{Liu2016a}.
Using the same technique, Shi~\textit{et~al.} found mismatch angles between MoS\textsubscript{2} and G, on top of a Cu foil, ranging from $\ang{-11}$ to $\ang{18}$, with a hexagonal flake size of about $\SI{1}{\mu m}$ \cite{shi2012}.
For CVD-grown flakes of $\SI{9}{\mu m}$, Lu~\textit{et~al.} found a mismatch with typical angles below $\ang{3}$ \cite{lu2015}.
Finally, using an exfoliation protocol, Adrian~\textit{et~al.} prepared multi-layered heterostructures and observed a misfit angle of $\ang{7.3}$ \cite{adrian2016}.
As well as different values for the observed mismatch angle, these studies offer different explanations for its origin.
Whereas some attribute the observed (mis)alignment to the vdW epitaxy accommodating the mismatch in lattice constants \cite{Liu2016a, shi2012}, others use the structural features of the underlying G and the edges \cite{lu2015} as an explanation.

In a recent theoretical work, Zhu~\textit{et~al.}~\cite{zhu2019} explained the orientational ordering of finite size homostructures, e.g. MoS$_2$ flakes on an MoS$_2$ substrate, using a purely geometrical argument: the lowest energy configuration is the one obtained by the roto-translation of the rigid flake which maximizes the number of locally commensurate regions.
Although this argument is solely based on geometry, it provides a satisfactory approximation for finite size systems but in the limit of infinite planes, i.e. for large enough flakes, commensurate regions equal incommensurate ones.
Therefore, in the limit of extended interfaces, other theoretical frameworks are needed.
In this contribution, we aim to explore the energy landscape originating from the rotational degree of freedom of edge-free, complex layered heterostructures and relate its fundamental origin to incommensurability and layer deformation at imposed angles.
This will provide guidance for the design of vdW heterostructures and the control of the twisting degree of freedom.
In order to make a more general point about the relative importance of different contributions in determining the total energy landscape, we focus on a specific but well-studied system, namely MoS$_2$/G.
This selected analysis shows the practical application of our argument and will also allow us to comment on the apparently contradictory experimental observations regarding this particular system.

\section*{Results and discussion}
In order to avoid finite-size effects and harvest information solely from the relaxation of the atoms in the layers, we implemented a protocol to build edge-free geometries.
The resulting supercells are simultaneously compatible with the lattice mismatch and a relative imposed angle between the lattices.
As a result, periodic boundary conditions (PBC) can be applied to such cells.
The starting interface geometry is described by a pair of 2D lattices defined by vectors $(l_{\rm{a}} \hat{\mathbf{a}}_1,l_{\rm{a}} \hat{\mathbf{a}}_2)$ and $(l_{\rm{b}}\hat{\mathbf{b}}_1, l_{\rm{b}} \hat{\mathbf{b}}_2)$, where $l_{\rm{a}}$ and $l_{\rm{b}}$ represent the lattice constants and the $\hat{b}_i$ vectors are rotated by an angle $\theta$ with respect to $\hat{a}_i$.
Two layers will be compatible if they satisfy the matching condition
$
	l_{\rm{a}} ( n_{1} \hat{\mathbf{a}}_{1} + n_{2} \hat{\mathbf{a}}_{2} ) = l_{\rm{b}} ( m_{1} \hat{\mathbf{b}}_{1} + m_{2} \hat{\mathbf{b}}_{2}  ) ,
$
where the integer numbers $n_1$, $n_2$, $m_1$, $m_2$ are supercell indices representing the repetition along each lattice vector.
In practice, for incommensurate lattice constants, the matching condition yielding PBC-compliant supercells can only be satisfied approximately, i.e. the lattice spacing of one of the two component needs to deviate from its equilibrium value.
Here, in order to obtain suitable structures with imposed angles between \ang{0} and \ang{60}, we accept supercells satisfying $|l'-l|<\SI{5e-7}{\AA}$.
We apply the strain to the MoS$_2$ layer, which leads to a maximum strain $\epsilon=\frac{l'-l}{l}$ within the same order of magnitude, four orders less than reported strains in other computational studies \cite{Wang2017,Wang2015,Ding2016}.
This protocol yields a set of supercells, each of which has a different number of atoms up to 343893, created according to the four supercell indices resulting in an unique twisting angle, satisfying the matching condition.
Details of this protocol and all the parameters of the supercells used are reported in the Supplementary Information (SI).

In these supercells, we distinguish intralayer and interlayer interatomic interactions, resulting in the following Hamiltonian
\begin{equation}
    \label{eq:hamil}
    H = H_\mathrm{L_1} + H_\mathrm{L_2} + H_\mathrm{L_{1}L_2}.
\end{equation}
The G layer is modelled with the REBO potential~\cite{Brenner2002}, $H_\mathrm{L_1}= H_\mathrm{C}^\mathrm{(REBO)}$, while the 3-body Stillinger-Weber (SW) potential~\cite{Ding2016} is used for MoS$_2$, $ H_\mathrm{L_2}= H_\mathrm{MoS_2}^\mathrm{(SW)}$.
Interlayer coupling is described by the Lennard-Jones (LJ) potential
\begin{align}
    \label{eq:lj_def}
    H_\mathrm{L_{1}L_2} &= H_\mathrm{C-Mo,C-S}^\mathrm{(LJ)}\nonumber \\
                        &=\sum_{\mathclap{\substack{i\in C \\ j\in Mo,S}}} 4 \epsilon_{ij} \left[ \left(\frac{\sigma_{ij}}{r}\right)^{12} - \left(\frac{\sigma_{ij}}{r}\right)^6 \right].
\end{align}
Since interlayer interactions are especially relevant for the aim of this work, we refined the values for the C-Mo and C-S interactions that can be found in Ref. \cite{Ding2016}.
As a reference set, we computed the  binding energy curves at DFT level using the Vienna \textit{Ab initio} Simulation Package (VASP) \cite{Kresse1993,Kresse1999} within the Projector Augmented-Wave (PAW) framework \cite{Blochl1994}.
The exchange-correlation potential is approximated using the PBE functional \cite{Perdew1996} and the vdW dispersion is described by the DFT-D2 method~\cite{Grimme2006}.
After this procedure, we are able to perform energy minimizations using the conjugate gradient algorithm available within the LAMMPS package \cite{Plimpton1995}.
More details about the fitting and minimization procedure can be found in the SI.

An approximate theory for the orientational ordering of an incommensurate interface was proposed by Novaco and McTague~\cite{novaco1977,mctague1979}.
Although developed in the context of epitaxial growth of noble gas layers on metal surfaces, it has been successfully applied to the behaviour of mesoscopic colloidal layers in optical lattices~\cite{Brazda2018a} and metal clusters adsorbed on G~\cite{Panizon2017a}.
The assumption of the Novaco-McTague (NM) theory is that two purely 2D systems linked via an interface may be divided into two separate components: a soft adsorbate layer, treated within the harmonic approximation, atop a rigid substrate.
This means that one of the intralayer terms in Eq.~\ref{eq:hamil} is substituted by its harmonic approximation, while the coordinates of the second layer are frozen at its initial values, $r_0$.
For example, considering G as the adsorbate and MoS$_2$ as the substrate yield a total Hamiltonian of the form
\begin{align}
H_\mathrm{NM} = \left. H_\mathrm{C}^\mathrm{(REBO)}\right|_\mathrm{harmonic} + \left. H_\mathrm{MoS_2}^\mathrm{(SW)}\right|_{r_0} +  H_\mathrm{C-MO,C-S}.
\end{align}
If the substrate and the adsorbate lattices are incommensurate due to a mismatch in lattice constants, the system is frustrated: the intralayer interactions within the adsorbate favor the intrinsic equilibrium lattice spacing, while the interactions with the substrate drive the atoms away from this spacing.
In the limit of long wavelength distortions, the NM theory predicts that the system can lower its energy by converting part of the longitudinal stress coming from the incommensurability into shear stress.
This yields a small misalignment angle between the two lattices given by
\begin{equation}
    \label{eq:NM_min}
    \cos\theta_\mathrm{NM} = \frac{1+ \rho^{2}(1+2\delta)}{\rho[2+\delta(1+\rho^{2})]},
\end{equation}
where $\rho=l_{\rm{substrate}}/l_{\rm{adsorbate}}$ is the mismatch ratio between the two lattices and $\delta=(c_{\rm{L}}/c_{\rm{T}})^{2}-1$, with $c_{\rm{T}}$ and $c_{\rm{L}}$ being the transverse and longitudinal sound velocities of the adsorbate, respectively.

The result of NM in Eq.~\ref{eq:NM_min} can be applied to our system by extracting the sound velocity of each single layer from the phonon dispersion, as reported in the SI.
There are two possible scenarios: G can be treated as the rigid substrate, while MoS$_2$ acts as a soft adsorbate, or vice versa.
In the first case, the theory predicts $\theta_\mathrm{NM}^\mathrm{MoS_2} = 8.0^{\circ}$, while if G is the adsorbate, the minimum-energy angle is $\theta_\mathrm{NM}^\mathrm{G} = 8.6^{\circ}$.
The prediction of the NM model can be verified by minimizing the total energy of the twisted geometries described above under suitable constraints.
We froze the atoms of the heterostructures in the direction perpendicular to the surface, i.e. the $z$ axis, effectively reducing the dimensionality of the system to 2D.
Furthermore, we also froze the atoms of the substrate layer in the in-plane directions $x$ and $y$, enforcing a fully rigid substrate.

As mentioned at the beginning of this section, the edge-free geometries used in this work inevitably retain a degree of stress resulting from the matching condition for the two lattices in order to be able to apply PBC.
This stress leads to a significant noise in the signal of the energy profile since the in-plane movements of the ions are small and directly affected by the imposed strain.
To overcome this problem and to obtain a clear signal in these simulations, the LJ-coupling strength between the MoS$_2$ and G layers was enhanced.
During the geometry optimization we set the LJ-parameters $\epsilon_{ij}$ in Eq.~\ref{eq:lj_def} to $\epsilon_{ij}' = 100\cdot \epsilon_{ij}$.
Next, the resulting energy profile is scaled back, as if simulated with the original value $\epsilon_{ij}' = \epsilon_{ij}$.
As is shown in the SI, this computational trick reduces the noise without affecting the actual physics of the problem.
\begin{figure}[]
  \centering
  \includegraphics[width=0.80\textwidth]{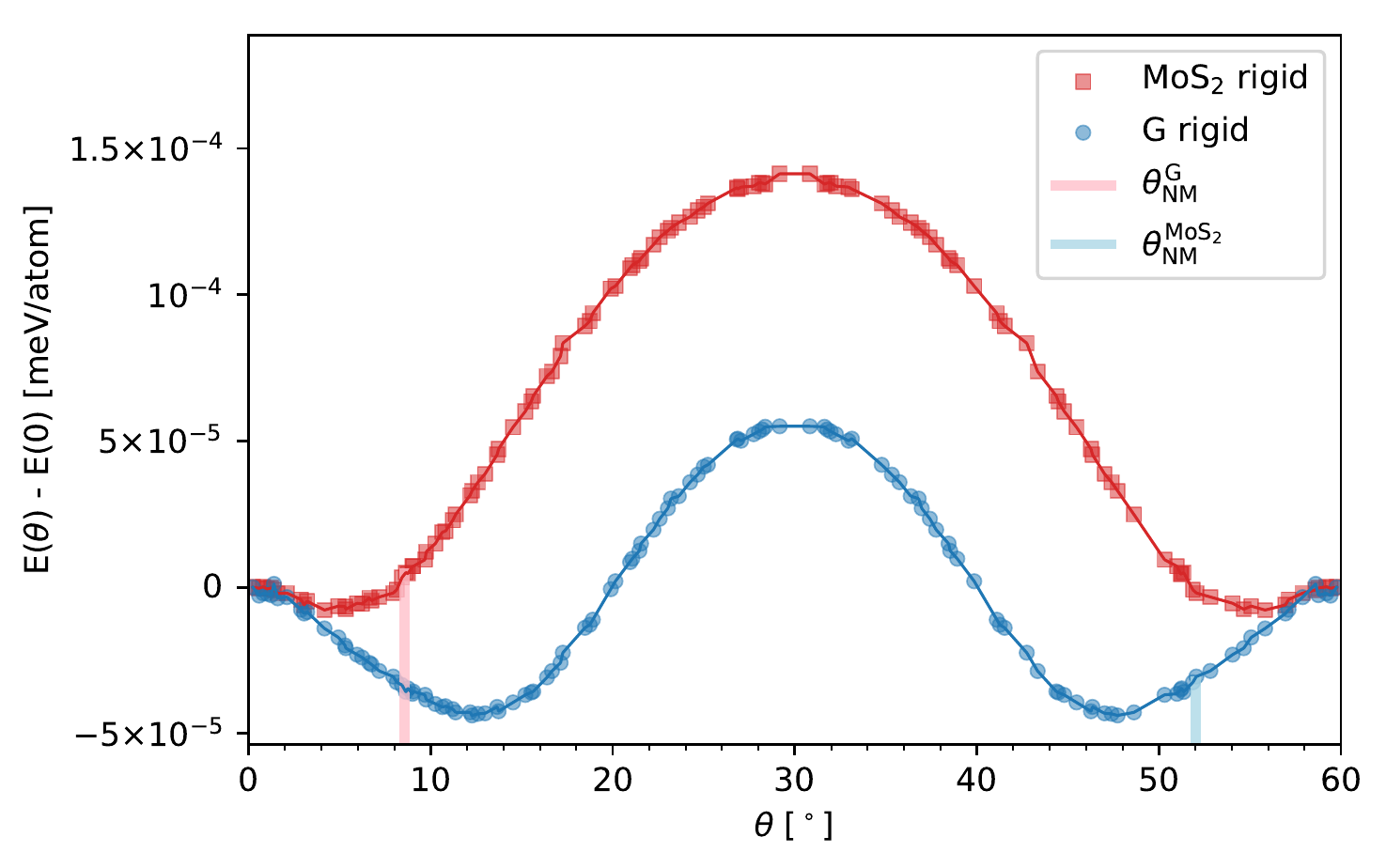}
  \caption{\label{fig:lj_novaco}
  Energy per atom $E(\theta)$, in meV, as a function of the imposed angle $\theta$ in degrees for different 2D models:
  red squares refer to flexible G on top of rigid MoS$_2$;
  blue circles refer to flexible MoS$_2$ on top of rigid G.
  The reference value of the energy scale is set to $E(0)$.
  Red and blue segments mark the minimum angle predicted by the NM theory for the first and second case, respectively.
  }
\end{figure}

Figure \ref{fig:lj_novaco} shows the optimized energy per atom, $E(\theta)$, of the bilayer system as a function of the angle, $\theta$, with respect to the energy of the aligned structures, $E(0)$.
The two curves refer to the following models: 2D-adsorbed G atop rigid MoS$_2$ (red) and 2D-adsorbed MoS$_2$ atop rigid G (blue).
Both cases reveal a minimum at a non-zero angle: for the adsorbed G case, the minimum is found at $\theta\approx\ang{6}$ while for the adsorbed MoS$_2$ case it is at $\theta\approx\ang{12}$.
The simulations show that the physics described by the approximation of Eq.~\ref{eq:NM_min} is still valid, i.e. a non-zero minimum angle is observed for both cases.
However, the absolute values of the predicted and observed angles are not in agreement.

A previous study~\cite{Guerra2017}, dealing with G and h-BN, showed that the NM model quantitatively describes the relaxation of the constrained system of these purely 2D materials.
Here, the NM theory captures the basics of the physics but is not able to describe satisfactorily the complex geometry of the bilayer system, especially in case of the flexible MoS$_2$ layer.
We attribute the poor prediction of the theory in our case to the internal 3D structure of the MoS$_2$ monolayer, which indeed is unaccounted for in the NM model.
This suggests that the NM theory is generally of limited utility for any bilayer comprising TMDs or other systems with a multi-atom thick single layer.

Indeed, the NM theory is even qualitatively inadequate if all degrees of freedom are considered, i.e. all atoms are free to move in the 3D space.
\begin{figure}[t]
    \centering
    \includegraphics[width=0.80\textwidth]{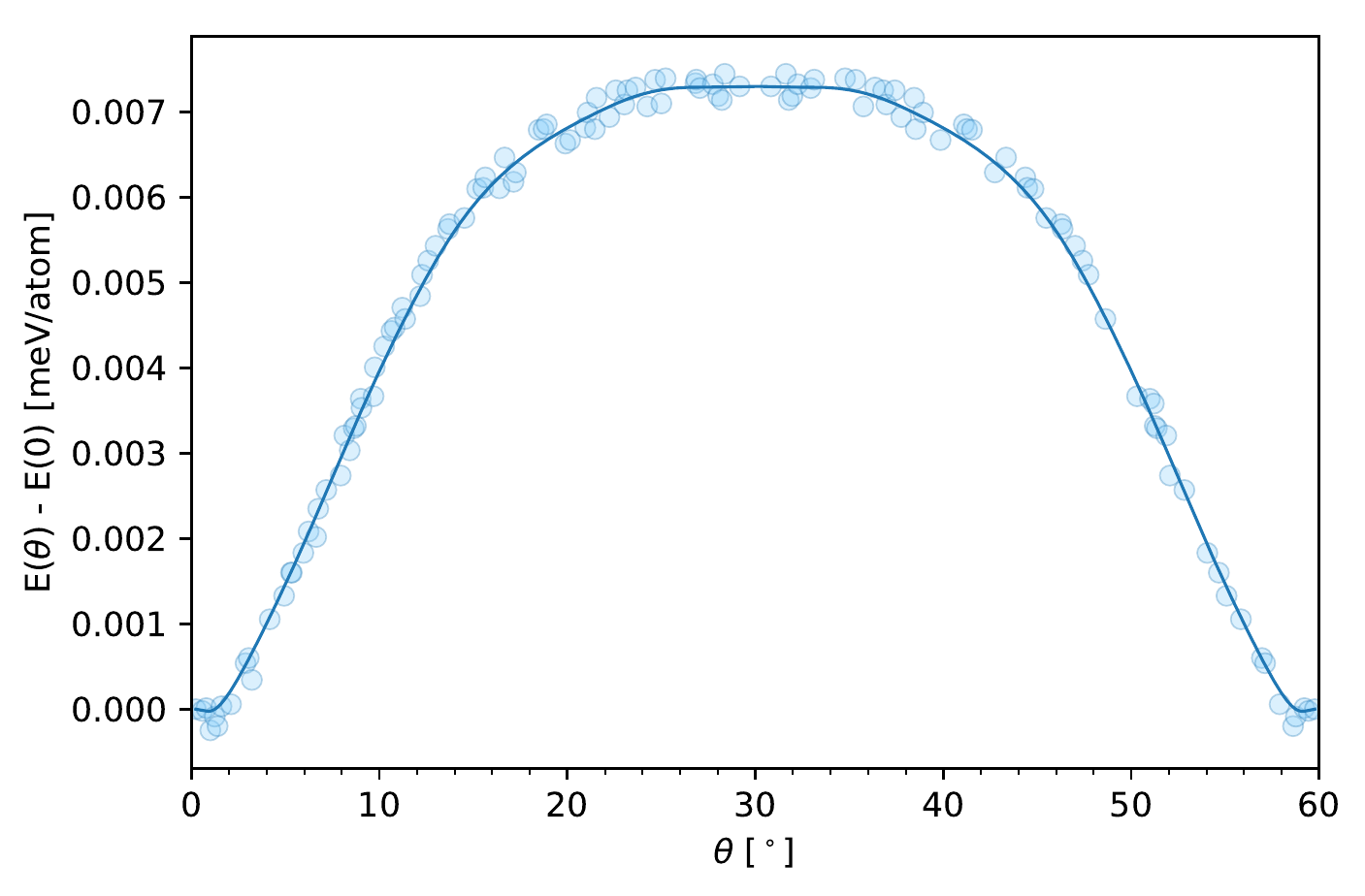}
    \caption{ \label{fig:en-force}
    Energy per atom $E(\theta)$, in meV, as a function of the imposed angle $\theta$.
    Each point in the energy landscape represents a distinct geometry at a different imposed angle and the blue line is a B\'ezier fit.
    The small oscillations at $\theta=\ang{0},\ang{60}$ are due to numerical noise in the energy simulations.
  }
\end{figure}
Figure \ref{fig:en-force} shows the energy per atom as a function of the angle of the system with no rigid substrate, but two soft, interacting layers.
Naturally, the LJ-coupling between the two layers has been restored to the values obtained from fitting against the DFT data to correctly reproduce interlayer forces.
The behaviour is both quantitatively and qualitatively different from the constrained system presented previously.
The introduction of the out-of-plane dimension ($z$) changes the response qualitatively.
The energy minima at non-zero angles have disappeared and $E(\theta)$ rises from the aligned cases (\ang{0},\ang{60}) with increasing mismatch angle up to \ang{30}.
From Fig.~\ref{fig:en-force}, one can thus deduce that at \SI{0}{K}, the fully flexible bilayer system will either be stable when aligned at $\ang{0}$ or $\ang{60}$, or mis-aligned at \ang{30}.

The NM theory does not hold when structural distortions perpendicular to the interface are allowed.
Our results indicate that these are important for MoS$_2$/G heterostructures and we have reason to believe that this is more widely the case.
The core of the NM argument is that the collective misalignment arises from the excitation of the transverse phonon branch in the $xy$ plane, which leads to a better interdigitation of the two lattices.
This excitation lowers the total energy of the system, as the transverse branch lies lower in energy than the longitudinal one.
As shown in Fig.~\ref{fig:phn} for G, the flexural band is flat near the $\Gamma$ point, i.e. the long-wave modulations perpendicular to the basal plane can occur essentially without an energy penalty.
If such distortions lead to a better interplay between the two layers, i.e. a gain in the interlayer coupling energy that is larger than the intralayer energy penalty from out-of-plane modulations, the system will lower its total energy.
Differently from the NM theory, the lowest-energy distortion in this scenario would not result in a misalignment between the components but in the formation of ripples keeping the locally commensurate zones at the equilibrium distance and pushing away incommensurate ones.

A tool to quantify the geometrical correspondence between lattices is the Moir\'e pattern.
The Moir\'e superlattice is a geometrical construction describing the interference between two lattices and can be used to identify zones of local commensuration between the two lattices.
The lattice parameter of the superlattice $L_{\rm{M}}$ is given by~\cite{mandelli2015}
\begin{equation}
    \label{eq:L_moire}
    L_{\rm{M}} = \frac{l_{\rm{G}}}{\sqrt{1+\rho^{-2} - 2 \rho^{-1}\cos\theta}}.
\end{equation}
\begin{figure}[t]
  \includegraphics[width=0.6\textwidth]{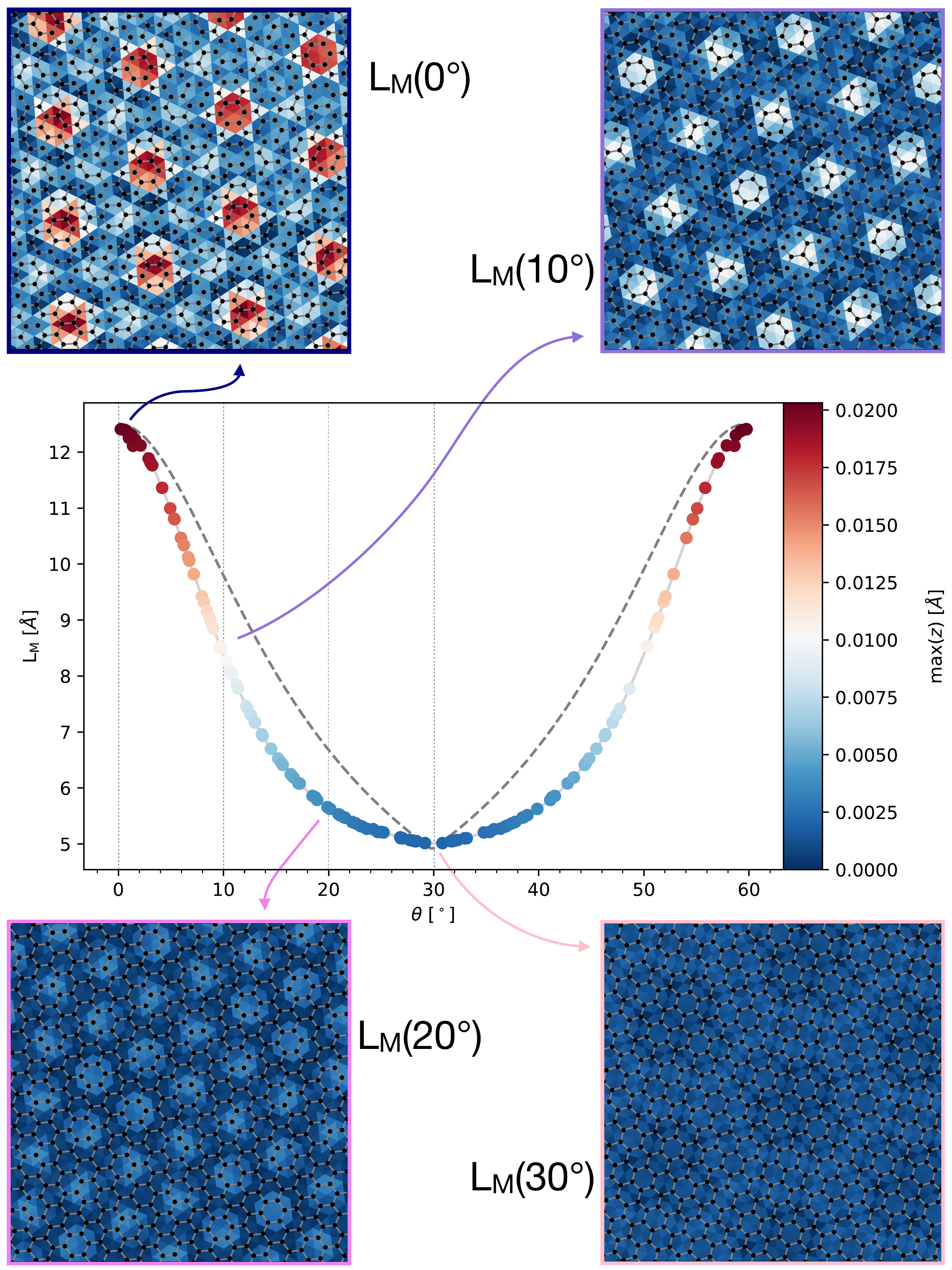}
  \caption{\label{fig:moire}
  Maximum $z$ displacement of C atoms (colored marks, right axis) and spacing of the Moir\'e pattern $L_{\rm{M}}$ (gray dashed line, left axis) as a function of $\theta$ for equivalent configurations at $\ang{0}$ and $\ang{60}$ rotating towards $\ang{30}$.
  The insets show the local distortion following the Moir\'e lattice in a square of sides $\SI{60}{\AA}$ at $\ang{0}$, $\ang{10}$, $\ang{20}$ and $\ang{30}$.
  The color of each triangle reports the $z$ coordinate of the corresponding C atom (black points) following the same color code as the main plot.
  For example, the Moir\'e pattern can be seen in the inset for $\mathrm{L_M}(\ang{0})$ as the lattice defined by the red regions.
  }
\end{figure}
Figure \ref{fig:moire} shows the correlation between the rippling in the $z$ dimension and the Moir\'e pattern in the G sheet.
At $\theta=0$, the Moir\'e spacing and the average displacement along $z$ are at a maximum and they both decrease as the misalignment increases.
This global parameter originates from the ripples of the carbon sheet, whose patterns follow perfectly the Moir\'e superlattice, as shown in the insets of Fig.~\ref{fig:moire} for selected values of $\theta$.
As $\theta$ increases, the length of the pattern shrinks with the displacement along $z$: at $\theta=\ang{30}$ the Moir\'e shrinks to a couple of unit cells and the monolayer remains basically flat.
\begin{figure}[t]
  \includegraphics[width=0.7\textwidth]{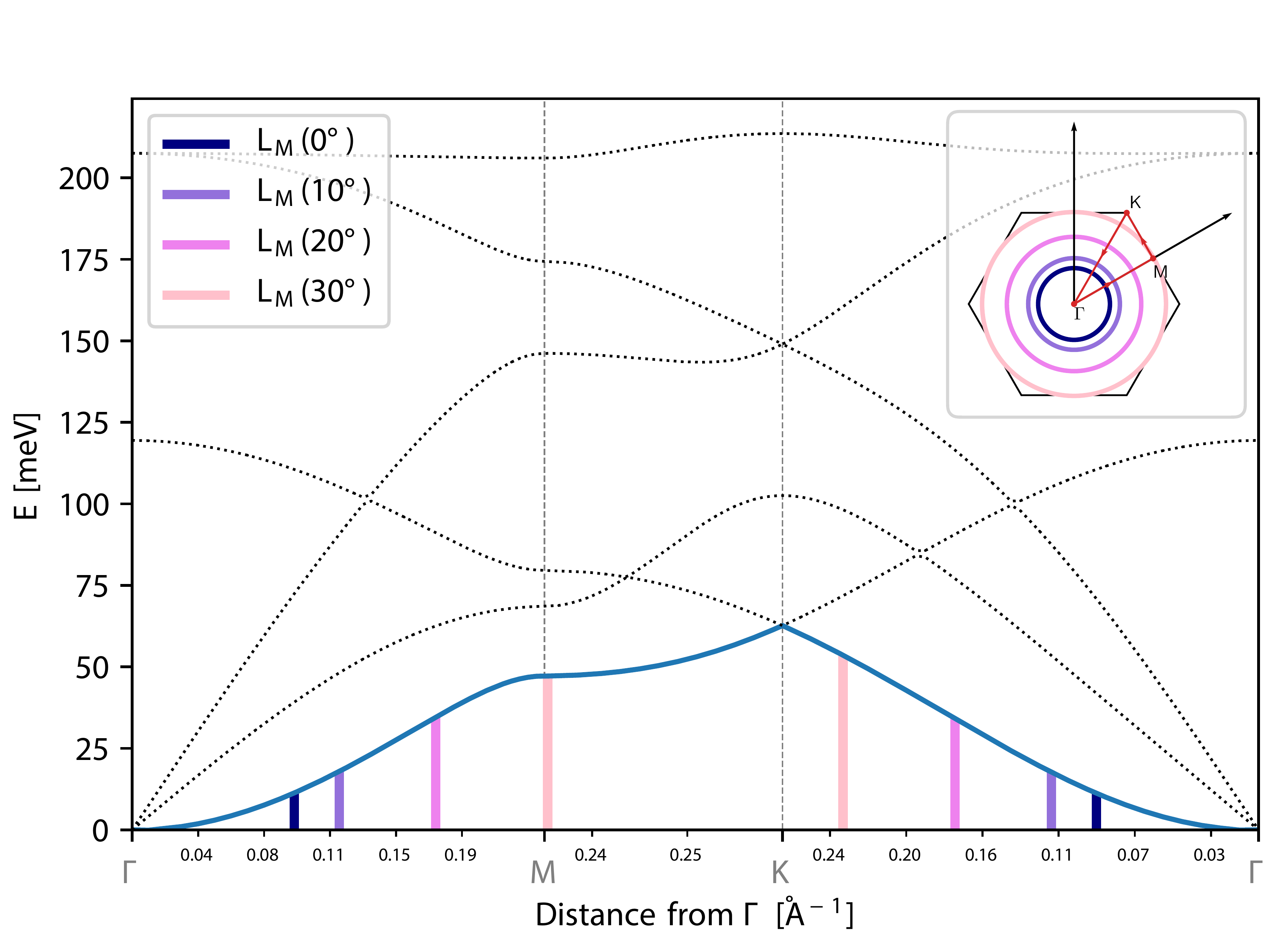}
  \caption{\label{fig:phn}
   Phonon band structure of the G monolayer.
   The $y$ axis reports the phonon energy, while the $x$ axis marks the distance from the origin along the path $\Gamma \to M \to K \to \Gamma$, shown in the top right inset and marked along the $x$ axis by gray dashed lines.
   The flexural branch is reported by a solid blue line while other branches are shown in dotted black lines.
   Colored segments along $x$ raising from $y=0$ to the flexural branch mark wavevectors matching the Moir\'e spacing $L_\mathrm{M}(\theta)$ for the geometries in the insets of Fig.~\ref{fig:moire}, as highlighted by the color-code.
   The Moir\'e wavelength in reciprocal space is computed as $k_\mathrm{M}=\frac{2}{\sqrt{3}L_\mathrm{M}}$ and the wavevectors matching it in the Brillouin zone are shown in the top-right inset following the color-code.
   }
\end{figure}
The connection between the geometries shown in Fig.~\ref{fig:moire} and the energy profile in Fig.~\ref{fig:en-force} can be understood in terms of the phonon dispersion of G.
This is reported in Fig.~\ref{fig:phn} and the wavevectors at which the acoustic flexural branch matches the Moir\'e spacing $L_\mathrm{M}$ at $\theta=\ang{0},\ang{10},\ang{20},\ang{30}$ are highlighted by vertical segments.
At small $\theta$, the spacing of the Moir\'e is around $\SI{12}{\AA}$, which is the distance between the locally commensurate patches.
As signaled by the dark-purple line in Fig.~\ref{fig:phn}, flexural phonon modes of this length in G are close enough to the flat region around $\Gamma$ and therefore energetically inexpensive.
This allows commensurate regions to stay at the equilibrium interlayer position while incommensurate ones are pushed away from each other perpendicular to the basal planes.
As $\theta$ increases to $\ang{30}$, $L_{\rm{M}}$ decreases and thus the distance between locally commensurate areas reduces.
As a result, the deformation needs to occur over a shorter distance and its energy cost therefore increases.
At $\theta=\ang{30}$, $L_{M}\approx\SI{5}{\AA}$, which is about 2 G unit cells.
As shown by the pink line in Fig.~\ref{fig:phn}, deformations of this length scale are described by phonons at the edges of the Brillouin zone and are energetically more expensive than the gain coming from the interdigitation with the substrate.
Therefore the G sheet remains flat, at the expense of the interlayer coupling, resulting in a higher total energy of the heterostructure compared to the aligned case.
To sum up, the unconstrained 3D heterostructure lowers its energy by out-of-plane distortions according to the Moir\'e pattern.
This is particularly convenient at $\theta=0$, where $L_{\rm{M}}\approx\SI{12}{\AA}$: here the flexural distortion is almost without any energy penalty and the system lowers its energy by improving the interdigitation between G and the MoS$_2$ layer.
As $\theta$ increases, $L_{\rm{M}}$ decreases and the cost of the ripples overtakes the gain in energy due to local commensuration, yielding flat G and an increased total energy.

We explored the stability of twisted vdW heterostructures.
Although often overlooked, this phenomenon is of particular importance in the emerging field of twistronics, as it can be a decisive factor in the real-life application of such systems.
The energy as a function of imposed angle determines whether a device is at risk of rotating away from a prepared angle even if it can be prepared in a meta-stable state.
Our analysis of MoS$_2$/G heterostructures helps to clarify the scattered experimental data.
We find a single global minimum at $\theta=\ang{0}$ and \ang{60}: i.e. only epitaxial stacking is expected for the system at $\SI{0}{K}$.
However, experiments always present defects or intrinsic friction that might result in the emergence of activation energies, potentially trapping a system in a meta-stable (or even unstable) state.
In the limit where such effects become negligible, i.e. activation energy approaching zero, one would only observe aligned and $\ang{30}$-rotated heterostructures, in agreement with the results of Liu~\textit{et~al.}~\cite{Liu2016a}.
A possible experiment to test the validity of our results would be a systematic repetition of the aforementioned experiments, focused upon reducing deviations resulting from working conditions, e.g. annealing temperature.
We expect that the results of such an effort will confirm our findings: with a high enough annealing temperature and large enough flakes of significant quality, the bilayer system should be found in the aligned configuration, with possibly some outlier around \ang{30}.
Finally, we explain the origin of the observed energy economy in terms of the interplay between flexural phonon modes of the pristine compounds and the Moir\'e superlattice.
This insight is general in nature and can be applied to all layered materials and heterostructures, serving as a design tool for twistronic devices.
Stiffness in the out-of-plane direction should be considered a critical property in the design of such devices.
Soft flexural phonon modes might be a lower energy route out of frustration than twisting, hindering the possibility of stable rotated configurations.
Furthermore, our results show the need for a novel theory of epitaxy for layered materials, incorporating the flexural branches ignored in the NM theory and taking into account all phonon wavelengths.
The insights presented here can serve as a starting point for developing such a theory of the epitaxial growth for vdW heterostructures.

\section*{Methods}
\subparagraph*{Classical MD}
All energy minimizations of the rotated heterostructures have been performed using molecular dynamics by means of the LAMMPS package \cite{Plimpton1995} using the conjugate gradient algorithm, where the energy tolerance was set to $\SI{1e-15}{}$. 
The REBO potential\cite{Brenner2002} was used for G, whereas an adapted version of the 3-body Stillinger-Weber (SW) potential\cite{Ding2016} was used for MoS\textsubscript{2}.
To model the vdW we used an interlayer LJ potential.
To obtain the explicit values of the parameters, we refined the values that can be found in Ref. \cite{Ding2016}, of which we provide an elaborate description below. 

\subparagraph*{DFT calculations}
Ab initio calculations used to re-parametrise the force field were carried out using DFT as implemented in the Vienna \textit{Ab initio} Simulation Package (VASP) \cite{Kresse1993,Kresse1999} within the Projector Augmented-Wave (PAW) framework \cite{Blochl1994}.
The exchange-correlation potential is approximated using the PBE functional \cite{Perdew1996} and the vdW dispersion is described by DFT-D2 method~\cite{Grimme2006}.
A plane wave cut-off of $\SI{800}{eV}$ is adopted and the Brillouin zone was sampled using a $13\times13\times1$ mesh. 

\subparagraph*{Phonon calculation}
Phonon bands were computed with the aid of Phonopy~\cite{Togo2015}, which was coupled to LAMMPS using phonoLAMMPS~\cite{carreras2019}.
In both cases the phonon dispersion was computed using the frozen method employing a 5x5x1 supercell.
\section*{Data Availability}
All parameters and protocol used  during this study are included in this published article (and its supplementary information files).
Original datasets are available from the corresponding author on reasonable request.

\section*{Acknowledgments}
The authors are thankful to E. Tosatti, A. Vanossi, D. Mandelli and  M. Liao for the helpful discussions.
This project has received funding from the European Union's Horizon2020 research and innovation programme under grant agreement No. 721642: SOLUTION.
The authors acknowledge the use of the IRIDIS High Performance Computing Facility, and associated support services at the University of Southampton, in the completion of this work.
This work was also supported by The Ministry of Education, Youth and Sports from the Large Infrastructures for Research, Experimental Development and Innovations project ``IT4Innovations National Supercomputing Center – LM2015070" and by the project Novel nanostructures for engineering applications No. CZ.02.1.01/0.0/0.0/16\_026/0008396.

\section*{Author Contributions}
AS and VEPC carried out all calculations and data analysis and conceptualize the study in the first place.
PN, DK supervised extensively the study and TP guided the project.
All authors contributed to the writing of this work. 
Finally, PN, DK and TP secured funding acquisition.

\section*{Additional information}
\subparagraph{Supplementary information} accompanies the paper on the npj Computational Materials website (DOI).
\subparagraph{Competing interests:} The authors declare no competing interests.

\section*{Supplementary Information}

In the following, we present in more details the methods and technical details used in this work.
%
%
In the first section, we describe the protocol to obtain the supercells of the heterostructures.
We also provide a table with the parameters of all structures used in this study.
In the second section, we explain the protocol used to refine the force field parameters, including the updated parameters.
In the third section, we prove that the enhancement of the LJ parameters does not alter the actual physics of the problem. 
In the fourth section, we discuss the phonon dispersion and how we extracted from it the quantities required as input for the NM approximation. 
Finally, in the fifth section, we further investigate the limits of the NM theory by modelling a constrained system midway between pure NM assumptions and a fully free bilayer.

\section{Supercells for twisted lattices}
Here, we explain the procedure to obtain the twisted lattices supercells.
Let $l_{\rm{a}}$ and $l_{\rm{b}}$ be the spacing of the Bravais lattices of layer a  and layer b, respectively and $\hat{\mathbf{a}}_{1}=\left( \begin{smallmatrix} 1 \\ 0 \end{smallmatrix} \right)$ be one of the primitive versors of the first lattice, aligned with the $x$ axis; the lattice with the desired periodicity is generated by a primitive vector $\mathbf{a}_{1}= l_{\rm{a}}\hat{\mathbf{a}}_{1}$. 
The matrix representing the discrete rotational symmetry of the lattice by an angle $\Omega=\pi/3$ is:
\begin{align}
\doubleunderline{R}_{\Omega}= 
\left(\begin{matrix}
	\cos\pi/3 & -\sin\pi/3 \\  
	\sin\pi/3 & \cos\pi/3 
\end{matrix} \right)
= 
\left( \begin{matrix}
	1/2 & -\sqrt{3}/3 \\  
	\sqrt{3}/3 & 1/2 
\end{matrix}\right) .
\end{align} 
Thus, the second versor defining the lattice is $\hat{\mathbf{a}}_{2}=\doubleunderline{R}_{\Omega}\hat{\mathbf{a}}_{1}$.
Since the second lattice, b, has the same symmetry but is rotated with respect to the first one by an angle $\theta$, versors defining it are
$
	(\hat{\mathbf{b}}_{1}, \hat{\mathbf{b}}_{2}) =
	 (\doubleunderline{R}_{\theta} \hat{\mathbf{a}}_{1}, ~ \doubleunderline{R}_{\Omega} \doubleunderline{R}_{\theta} \hat{\mathbf{a}}_{1} ) 
$
where 
\begin{align}
    \doubleunderline{R}_{\theta}=
    \left( 
        \begin{matrix} \cos\theta & -\sin\theta \\                     \sin\theta & \cos\theta \end{matrix} 
    \right)
\end{align}
describes the misalignment between the lattices.
A heterostructure supercell will be compatible with both periodicities if the individual lattice cells match exactly at the edges, in other words, if the following matching condition is satisfied
\begin{equation}
    \label{eq:matching_condition}
    l_{\rm{a}} ( n_{1} \hat{\mathbf{a}}_{1} + n_{2} \hat{\mathbf{a}}_{2} ) = l_{\rm{b}} ( m_{1} \hat{\mathbf{b}}_{1} + m_{2} \hat{\mathbf{b}}_{2}  ),
\end{equation}
where $n_1, n_2, m_1, m_2$ represent the repetition along the corresponding versor of the unit cell of the first and second lattice, respectively.
This condition can be rewritten with a matrix formalism to:
\begin{align} 
	\frac{l_{\rm{a}}}{l_{\rm{b}}} ( n_{1} \hat{\mathbf{a}}_{1} + n_{2} \doubleunderline{R}_{\Omega}\cdot \hat{\mathbf{a}}_{1} ) = 
	m_{1}  \doubleunderline{R}_{\theta}\cdot \hat{\mathbf{a}}_{1}  + m_{2} \doubleunderline{R}_{\Omega}\cdot \doubleunderline{R}_{\theta}\cdot \hat{\mathbf{a}}_{1}  \nonumber
	\\
	\rho \left( \begin{array}{cc} \mathbb{I} & \doubleunderline{R}_{\Omega}\end{array}\right) \cdot \left( \begin{array}{c} n_{1} \\ n_{2} \end{array} \right)  = 
	\left( \begin{array}{cc} \mathbb{I} & \doubleunderline{R}_{\Omega}\end{array} \right) \cdot \left( \begin{array}{c} m_{1} \\ m_{2} \end{array} \right)  \cdot \doubleunderline{R}_{\theta}  \label{eq:match_vec}
\end{align}
where $\mathbb{I}$ is the identity matrix, we used the definition of the lattice vectors, introduced the mismatch ratio $\rho=l_{\rm{a}}/l_{\rm{b}}$, grouped the matrices and the indexes in vectors and simplified $\hat{\mathbf{a}}_{1}$ from both sides.

Albeit that the mismatch ratio of a system is fixed by the equilibrium values of the lattice parameters, it would be impractical to approximate a real number using integers, as the size of the supercells would easily exceed our computational capabilities.
We follow the reverse procedure: given the four indexes $\{m_{i}, n_{i}\}_{i=1,2}$, we can invert the system and find the mismatch ratio $\rho$ and the misalignment angle $\theta$ that satisfy the matching condition of Eq. \ref{eq:matching_condition}.
This means that now $\{m_{i}, n_{i}\}_{i=1,2}$ are fixed parameters of Eq. \ref{eq:matching_condition} while $\rho$ is a variable, along with $\theta$.
Next, we find an expression for $\rho$ and $\theta$ in terms of $\{m_{i}, n_{i}\}_{i=1,2}$ that satisfies Eq. \ref{eq:matching_condition}.
In the following paragraph, we address the problem of selecting sets of indices whose corresponding $\rho$, is close enough the real value fixed by the system $\rho_0$.
We solve equation \eqref{eq:match_vec} for the matrix $\doubleunderline{R}_{\theta}$ and for $\rho$ under the constraint that  $\doubleunderline{R}_{\theta}$ is a rotation matrix, namely:
\begin{align}
\label{eq:match_system} 
\begin{cases}
	 & \doubleunderline{R}_{\theta}  = \rho \left(  m_{1} \mathbb{I} + m_{2} \doubleunderline{R}_{\Omega}\right)^{-1} \left( \begin{array}{cc} \mathbb{I} & \doubleunderline{R}_{\Omega}\end{array}\right) \cdot \left( \begin{array}{c} n_{1} \\ n_{2} \end{array} \right) 
	 \\
	 & \det \doubleunderline{R}_{\theta}= 1.
	 \end{cases}
\end{align}
%
The first line in \cref{eq:match_system} is readily solved by
\begin{widetext}
\begin{align}
	 \doubleunderline{R}_{\theta} &= \frac{\rho}{N_{\rm{b}}} 
		\left( \begin{matrix}
		m_{1} n_{1} + m_{2}n_{2}  + 1/2 (m_{1}n_{2} +m_{2} n_{1} )   	&  	- \sqrt{3}/2 ( m_{1} n_{2}  -  m_{2} n_{1}  ) \\   
		\sqrt{3}/2 ( m_{1} n_{2}  -  m_{2} n_{1}  ) 							& 	m_{1} n_{1} + m_{2}n_{2}  + 1/2 (m_{1}n_{2} +m_{2} n_{1})
		\end{matrix}\right) \nonumber \\
	 &= \frac{\rho}{N_{\rm{b}}} \doubleunderline{A} \label{eq:rsol} ,
\end{align}
\end{widetext}
where $N_{\rm{b}}=m_{1}^{2} + m_{2}^{2} + m_{1}m_{2}$ is the number of Bravais lattice points in the b lattice~\footnote{An equivalent relation holds for the other lattice $N_{\rm{a}}=n_{1}^{2} + n_{2}^{2} + n_{1}n_{2}$} and  $\doubleunderline{A}$, implicitly defined in the last step, is a shorthand for the matrix of known coefficients.
Substituting \cref{eq:rsol} into the second line of \cref{eq:match_system} yields an expression for $\rho$: $
    \det \doubleunderline{R}_{\theta} = \frac{\rho^{2}}{N_{\rm{b}}^{2}} \det \doubleunderline{A}
    = 1$.
Substituting this back into \cref{eq:rsol} gives us the solution of $(\rho, \theta)$ of \cref{eq:matching_condition} at chosen $\{m_{i}, n_{i}\}_{i=1,2}$: 
\begin{align}
\begin{cases}
	\rho &= \frac{N_{\rm{b}}}{\sqrt{\det \doubleunderline{A}}} \label{eq:rhosol}
	\\
	\theta &= (\doubleunderline{R}_{\theta})_{11} =  \arccos \left( \frac{1}{\sqrt{\det \doubleunderline{A}}} A_{11}\right) 
\end{cases}
\end{align}

%
Finally, the first vector of the supercell is given by the one of the members of the equality in \cref{eq:matching_condition} and the second is obtained by symmetry, namely
\begin{align}
	&\mathbf{C_1}= l_{\rm{a}} ( n_1 \hat{\mathbf{a}}_1 + n_2 \hat{\mathbf{a}}_2 ) \\
	&\mathbf{C_2}= \doubleunderline{R}_\Omega \cdot \mathbf{C_1}= -l_{\rm{a}} n_1 \hat{\mathbf{a}}_1 + l_{\rm{a}}(n_1+n_2)\hat{\mathbf{a}}_2 .
\end{align}

In order to obtain a system with the desired misalignment $\theta$ and a $\rho$ that is an acceptable approximation of the equilibrium mismatch $\rho_0$, we consider all combinations of integers $n_i,~m_i$ within the range $(-200,200)$ and select the supercells which satisfy $\theta \in [\ang{0},\ang{60}]$ and a mismatch $\rho$ satisfying $|\rho-\rho_0|\leq \SI{1e-7}{}$.
We then bin the resulting supercells with a spacing of $\Delta\theta=\ang{0.01}$ and select the cell with the smaller number of Bravais point within each bin.
Note that this procedure does not guarantee that the resulting supercell will be evenly spaced according to the mismatch angle.

The indices defining the supercells used in this work for the MoS$_2$/G heterostructures are reported in Tab. \ref{tab:rotsc_param}, along with the misalignment angle, $\rho-\rho_0$ and number atoms in each layer.
For this system $\rho_0=l_{\rm{G}}/l_{\rm{MoS_2}}=\SI{2.4601878}{\AA}/\SI{3.0936827}{\AA}=0.7952295$, the number of atoms in each lattice is given by the number of Bravais lattice points times the number of atoms in the crystal basis, i.e. $N_{\rm{tot}}=N_{\rm{Bravais}} \cdot n_{\rm{basis}}$ with $n_{\rm{basis}}$ is 2 and 3 for G and MoS$_2$, respectively.
In creating the supercell, the strain due to the approximated mismatch $\rho$ is applied to MoS$_2$, leading to the strain $\epsilon \approx \SI{1e-7}{}$ as mentioned in the main text.
%
\begin{longtable}{|m{1.8cm}|m{1.7cm}|m{1.7cm}|m{1.7cm}|m{1.7cm}|m{2.3cm}|m{2.1cm}|m{2.1cm}| }
\hline
$\theta$[$^\circ$]	&	$n_1$	&	$n_2$	&	$m_1$	&	$m_2$	&	$\rho-\rho_0$	&	$N_{\rm{G}}$	&	$N_{\rm{MoS_2}} $			\\
\hline
0.23	&	-135	&	-104	&	-108	&	-82	&	9.8e-08	&	86162	&	81732	\\
0.58	&	-192	&	61	&	-153	&	50	&	8.1e-08	&	57744	&	54774	\\
0.79	&	-184	&	-41	&	-148	&	-30	&	9.8e-08	&	86162	&	81732	\\
1.01	&	-113	&	-182	&	-86	&	-148	&	-6.2e-08	&	132916	&	126081	\\
1.24	&	-176	&	-59	&	-137	&	-51	&	4.3e-09	&	89682	&	85071	\\
1.39	&	-109	&	39	&	-87	&	33	&	-5.7e-08	&	18300	&	17361	\\
1.60	&	-141	&	-128	&	-107	&	-107	&	2.6e-08	&	108626	&	103038	\\
2.11	&	-138	&	-99	&	-104	&	-85	&	-2.2e-08	&	85012	&	80640	\\
2.88	&	-184	&	-41	&	-140	&	-42	&	9.8e-08	&	86162	&	81732	\\
3.05	&	-118	&	12	&	-96	&	15	&	9e-08	&	25302	&	24000	\\
3.21	&	-182	&	-113	&	-134	&	-102	&	-6.2e-08	&	132916	&	126081	\\
4.17	&	-184	&	33	&	-142	&	15	&	8.1e-08	&	57744	&	54774	\\
4.93	&	-191	&	-48	&	-140	&	-55	&	3.5e-08	&	95904	&	90972	\\
5.30	&	-192	&	61	&	-155	&	62	&	8.1e-08	&	57744	&	54774	\\
5.34	&	-185	&	63	&	-149	&	63	&	4.6e-08	&	53076	&	50349	\\
5.95	&	-164	&	-41	&	-118	&	-50	&	1.5e-08	&	70600	&	66969	\\
6.23	&	-164	&	-115	&	-110	&	-113	&	6.2e-08	&	117960	&	111894	\\
6.64	&	-123	&	-185	&	-71	&	-169	&	-8.8e-08	&	144216	&	136800	\\
6.75	&	-164	&	-145	&	-154	&	-89	&	4.4e-08	&	143402	&	136029	\\
7.18	&	-152	&	-99	&	-140	&	-55	&	3.5e-08	&	95906	&	90972	\\
7.94	&	-172	&	169	&	-146	&	122	&	-8.3e-08	&	58152	&	55164	\\
8.14	&	-118	&	12	&	-99	&	24	&	9e-08	&	25302	&	24000	\\
8.42	&	-163	&	-125	&	-156	&	-68	&	-9.6e-08	&	125136	&	118704	\\
8.64	&	-145	&	145	&	-104	&	124	&	1.2e-08	&	42048	&	39885	\\
8.75	&	-138	&	-99	&	-85	&	-104	&	-2.2e-08	&	85012	&	80640	\\
9.01	&	-184	&	33	&	-153	&	50	&	8.1e-08	&	57744	&	54774	\\
9.04	&	-176	&	-59	&	-117	&	-76	&	4.3e-09	&	89682	&	85071	\\
9.69	&	-172	&	169	&	-122	&	146	&	-8.3e-08	&	58152	&	55161	\\
9.75	&	-181	&	-95	&	-113	&	-110	&	6.2e-08	&	117960	&	111894	\\
10.25	&	-169	&	-174	&	-90	&	-178	&	7e-08	&	176484	&	167409	\\
10.64	&	-181	&	-95	&	-110	&	-113	&	6.2e-08	&	117960	&	111894	\\
10.81	&	-164	&	-145	&	-89	&	-154	&	4.4e-08	&	143402	&	136029	\\
11.22	&	-174	&	-169	&	-90	&	-178	&	7e-08	&	176484	&	167409	\\
11.37	&	-170	&	71	&	-130	&	31	&	-4.5e-08	&	43740	&	41493	\\
12.17	&	-113	&	-182	&	-134	&	-102	&	-6.2e-08	&	132916	&	126081	\\
12.27	&	-185	&	63	&	-138	&	19	&	4.6e-08	&	53076	&	50349	\\
12.60	&	-192	&	61	&	-142	&	15	&	8.1e-08	&	57744	&	54774	\\
12.99	&	-135	&	-104	&	-140	&	-42	&	9.8e-08	&	86162	&	81732	\\
13.65	&	-118	&	12	&	-81	&	-15	&	9e-08	&	25302	&	24000	\\
13.73	&	-184	&	33	&	-155	&	62	&	8.1e-08	&	57744	&	54774	\\
14.52	&	-144	&	-98	&	-150	&	-31	&	1.7e-10	&	88904	&	84333	\\
15.20	&	-184	&	33	&	-127	&	-15	&	8.1e-08	&	57744	&	54774	\\
15.53	&	-185	&	63	&	-149	&	86	&	4.6e-08	&	53076	&	50349	\\
15.63	&	-184	&	-41	&	-108	&	-82	&	9.8e-08	&	86162	&	81732	\\
16.39	&	-182	&	-113	&	-86	&	-148	&	-6.2e-08	&	132916	&	126081	\\
16.66	&	-135	&	-104	&	-148	&	-30	&	9.8e-08	&	86162	&	81732	\\
17.14	&	-163	&	-125	&	-68	&	-156	&	-9.6e-08	&	125136	&	118704	\\
17.26	&	-109	&	39	&	-87	&	54	&	-5.7e-08	&	18300	&	17358	\\
18.47	&	-192	&	61	&	-155	&	93	&	8.1e-08	&	57744	&	54774	\\
18.74	&	-118	&	12	&	-75	&	-24	&	9e-08	&	25302	&	24000	\\
18.91	&	-41	&	-184	&	30	&	-178	&	9.8e-08	&	86162	&	81729	\\
19.89	&	-185	&	-123	&	-71	&	-169	&	-8.8e-08	&	144216	&	136800	\\
20.15	&	-184	&	-23	&	-101	&	-79	&	-7.1e-08	&	77232	&	73263	\\
20.95	&	-187	&	-101	&	-75	&	-153	&	-5.7e-08	&	128114	&	121527	\\
21.08	&	-152	&	-99	&	-55	&	-140	&	3.5e-08	&	95906	&	90972	\\
21.47	&	-182	&	-23	&	-173	&	48	&	-8.3e-08	&	75678	&	71784	\\
21.55	&	-135	&	-104	&	-42	&	-140	&	9.8e-08	&	86162	&	81732	\\
22.24	&	-170	&	71	&	-130	&	99	&	-4.5e-08	&	43740	&	41493	\\
22.58	&	-184	&	-41	&	-182	&	42	&	9.8e-08	&	86162	&	81732	\\
23.03	&	-176	&	-59	&	-76	&	-117	&	4.3e-09	&	89682	&	85071	\\
23.20	&	-192	&	61	&	-153	&	103	&	8.1e-08	&	57744	&	54774	\\
23.63	&	-192	&	61	&	-127	&	-15	&	8.1e-08	&	57744	&	54774	\\
24.25	&	-148	&	-132	&	-185	&	-15	&	-4.9e-08	&	117728	&	111675	\\
24.67	&	-184	&	-41	&	-82	&	-108	&	9.8e-08	&	86162	&	81732	\\
25.00	&	-113	&	-182	&	-174	&	-52	&	-6.2e-08	&	132916	&	126081	\\
25.23	&	-135	&	-104	&	-30	&	-148	&	9.8e-08	&	86162	&	81732	\\
26.82	&	-191	&	-48	&	-195	&	55	&	3.5e-08	&	95904	&	90972	\\
26.86	&	-185	&	63	&	-119	&	-19	&	4.6e-08	&	53076	&	50349	\\
27.05	&	-144	&	-98	&	-31	&	-150	&	1.7e-10	&	88904	&	84333	\\
27.74	&	-164	&	-41	&	-168	&	50	&	1.5e-08	&	70600	&	66969	\\
28.03	&	-148	&	-132	&	-15	&	-185	&	-4.9e-08	&	117728	&	111675	\\
28.22	&	-184	&	-23	&	-79	&	-101	&	-7.1e-08	&	77232	&	73263	\\
28.37	&	-184	&	33	&	-103	&	-50	&	8.1e-08	&	57744	&	54774	\\
29.17	&	-176	&	-59	&	-188	&	51	&	4.3e-09	&	89682	&	85071	\\
30.83	&	-176	&	-59	&	-51	&	-137	&	4.3e-09	&	89682	&	85071	\\
31.63	&	-184	&	33	&	-153	&	103	&	8.1e-08	&	57744	&	54774	\\
31.78	&	-23	&	-184	&	79	&	-180	&	-7.1e-08	&	77232	&	73260	\\
31.97	&	-148	&	-132	&	-200	&	15	&	-4.9e-08	&	117728	&	111675	\\
32.26	&	-164	&	-41	&	-50	&	-118	&	1.5e-08	&	70600	&	66969	\\
32.95	&	-144	&	-98	&	-181	&	31	&	1.7e-10	&	88904	&	84333	\\
33.14	&	-185	&	63	&	-138	&	119	&	4.6e-08	&	53076	&	50349	\\
34.77	&	-104	&	-135	&	30	&	-178	&	9.8e-08	&	86162	&	81729	\\
35.33	&	-41	&	-184	&	82	&	-190	&	9.8e-08	&	86162	&	81729	\\
35.75	&	-148	&	-132	&	15	&	-200	&	-4.9e-08	&	117728	&	111675	\\
36.37	&	-192	&	61	&	-142	&	127	&	8.1e-08	&	57744	&	54774	\\
36.80	&	-192	&	61	&	-103	&	-50	&	8.1e-08	&	57744	&	54774	\\
36.97	&	-59	&	-176	&	76	&	-193	&	4.3e-09	&	89682	&	85068	\\
37.42	&	-184	&	-41	&	-42	&	-140	&	9.8e-08	&	86162	&	81732	\\
37.76	&	-170	&	71	&	-99	&	-31	&	-4.5e-08	&	43740	&	41493	\\
38.45	&	-135	&	-104	&	-182	&	42	&	9.8e-08	&	86162	&	81732	\\
38.53	&	-23	&	-182	&	-125	&	-48	&	-8.3e-08	&	75678	&	71784	\\
38.92	&	-152	&	-99	&	-195	&	55	&	3.5e-08	&	95906	&	90972	\\
39.85	&	-23	&	-184	&	101	&	-180	&	-7.1e-08	&	77232	&	73260	\\
41.09	&	-184	&	-41	&	-30	&	-148	&	9.8e-08	&	86162	&	81732	\\
41.26	&	-118	&	12	&	-99	&	75	&	9e-08	&	25302	&	24000	\\
41.53	&	-192	&	61	&	-93	&	-62	&	8.1e-08	&	57744	&	54774	\\
42.74	&	-109	&	39	&	-54	&	-33	&	-5.7e-08	&	18300	&	17361	\\
43.34	&	-135	&	-104	&	30	&	-178	&	9.8e-08	&	86162	&	81729	\\
44.37	&	-41	&	-184	&	108	&	-190	&	9.8e-08	&	86162	&	81729	\\
44.47	&	-185	&	63	&	-86	&	-63	&	4.6e-08	&	53076	&	50349	\\
44.80	&	-184	&	33	&	-142	&	127	&	8.1e-08	&	57744	&	54774	\\
45.48	&	-144	&	-98	&	31	&	-181	&	1.7e-10	&	88904	&	84333	\\
46.27	&	-184	&	33	&	-62	&	-93	&	8.1e-08	&	57744	&	54774	\\
46.35	&	-118	&	12	&	-96	&	81	&	9e-08	&	25302	&	24000	\\
47.01	&	-135	&	-104	&	42	&	-182	&	9.8e-08	&	86162	&	81732	\\
47.40	&	-192	&	61	&	-127	&	142	&	8.1e-08	&	57744	&	54774	\\
47.73	&	-185	&	63	&	-119	&	138	&	4.6e-08	&	53076	&	50349	\\
48.63	&	-170	&	71	&	-99	&	130	&	-4.5e-08	&	43740	&	41493	\\
50.31	&	-172	&	169	&	-146	&	24	&	-8.3e-08	&	58152	&	55164	\\
50.99	&	-184	&	33	&	-50	&	-103	&	8.1e-08	&	57744	&	54774	\\
51.20	&	-104	&	-135	&	82	&	-190	&	9.8e-08	&	86162	&	81729	\\
51.25	&	-138	&	-99	&	-189	&	85	&	-2.2e-08	&	85012	&	80640	\\
51.36	&	-145	&	145	&	-124	&	20	&	1.2e-08	&	42048	&	39885	\\
51.86	&	-118	&	12	&	-24	&	-75	&	9e-08	&	25302	&	24000	\\
52.06	&	-172	&	169	&	-24	&	146	&	-8.3e-08	&	58152	&	55161	\\
52.82	&	-152	&	-99	&	55	&	-195	&	3.5e-08	&	95906	&	90972	\\
54.05	&	-164	&	-41	&	-168	&	118	&	1.5e-08	&	70600	&	66969	\\
54.66	&	-185	&	63	&	-63	&	-86	&	4.6e-08	&	53076	&	50349	\\
55.07	&	-191	&	-48	&	-195	&	140	&	3.5e-08	&	95904	&	90972	\\
55.83	&	-184	&	33	&	-127	&	142	&	8.1e-08	&	57744	&	54774	\\
56.95	&	-118	&	12	&	-15	&	-81	&	9e-08	&	25302	&	24000	\\
57.12	&	-184	&	-41	&	-182	&	140	&	9.8e-08	&	86162	&	81732	\\
57.89	&	-138	&	-99	&	-189	&	104	&	-2.2e-08	&	85012	&	80640	\\
58.61	&	-109	&	39	&	-33	&	-54	&	-5.7e-08	&	18300	&	17358	\\
58.76	&	-176	&	-59	&	-188	&	137	&	4.3e-09	&	89682	&	85071	\\
59.21	&	-184	&	-41	&	30	&	-178	&	9.8e-08	&	86162	&	81729	\\
59.42	&	-192	&	61	&	-50	&	-103	&	8.1e-08	&	57744	&	54774	\\
59.77	&	-135	&	-104	&	82	&	-190	&	9.8e-08	&	86162	&	81729	\\
\hline
\caption{\label{tab:rotsc_param}
Parameters of the rotated supercells used in this work.
}
\end{longtable}

\section{Force Field refinement}
\begin{figure}
  \includegraphics[width=0.89\textwidth]{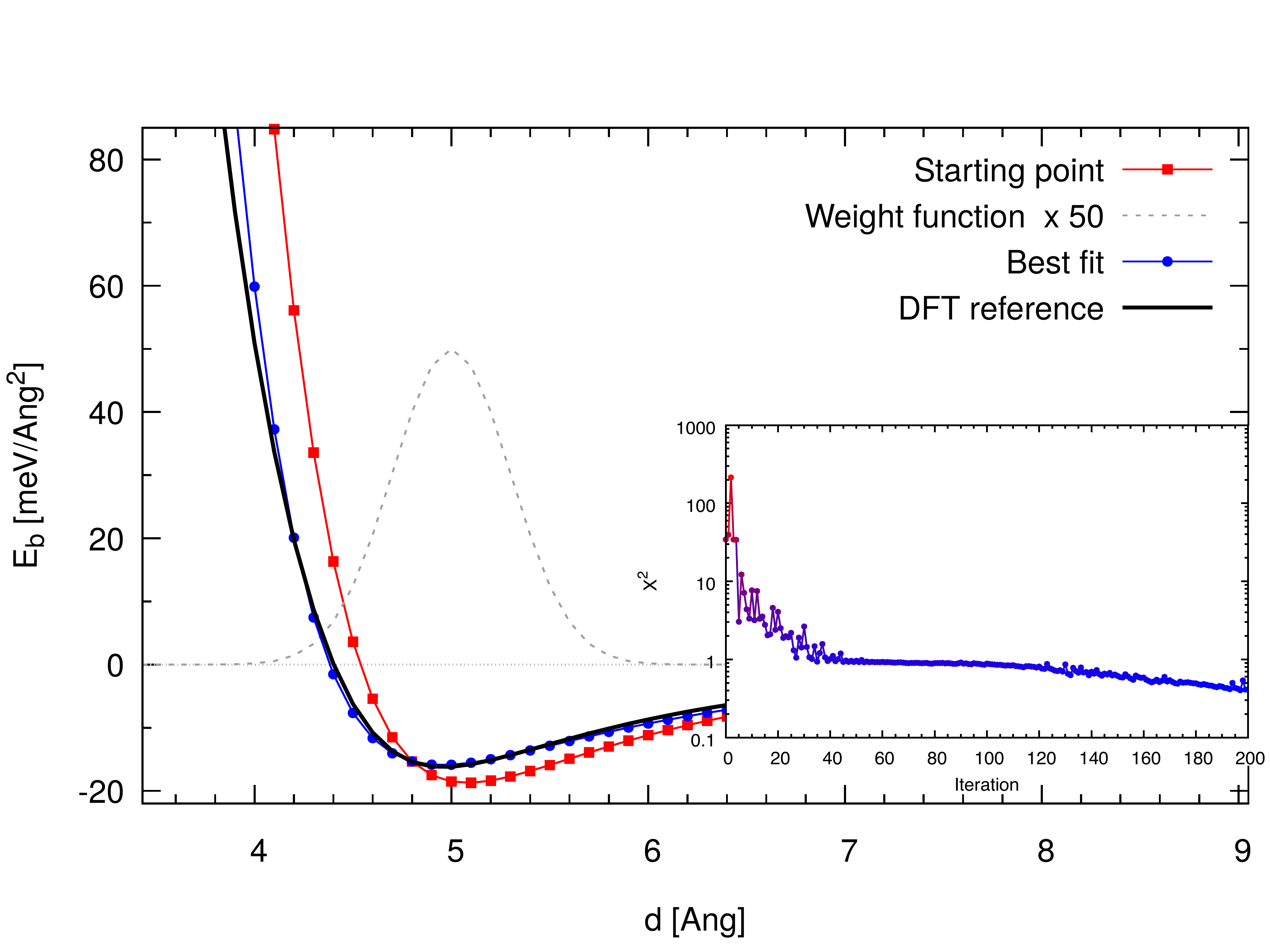}
  \caption{\label{fig:optim}
    Refining of LJ-parameters. Plotted are the energy per $\AA^2$ against the interlayer distance in $\AA$. The black line is the reference binding energy obtained through DFT, whereas the red and the blue line are the starting and final binding energies obtained through LAMMPS, respectively.The dashed line represents the weight function around the energy minimum enhanced by a factor of 50, as guide for the eye.
    The inset shows the goal function $\chi^2$ versus the number of the iterations of the optimization algorithm.
   }
\end{figure}

As part of this study, we refined the LJ coupling parameters for the heterostructure of MoS$_2$ and G. 
As a starting point, we took the parameters provided in Ref. \cite{Ding2016}.
However, our preliminary results revealed some significant discrepancies, both quantitatively as well as qualitatively, with the results obtained via accurate DFT calculations.
This motivated us to perform a recalibration of the parameters.
In order to do so, we applied the Simplex algorithm \cite{Nelder1965} as implemented in SciPy \cite{Oliphant2007scipy}.
This algorithm samples the N-dimensional (N$=$number of LJ parameters) phase space using a convex polygon.
This algorithm acts on the following goal function:
\begin{equation}
    \label{eq:simplex}
    \chi^2[f_L]=\frac{1}{W}\int_0^{\infty}
    |f_{DFT}(r)-f_L(r)|^2w(r)dr
\end{equation}
which is a squared distance combined with a weight function $w(r)$.
The function $f_{DFT}$ is the reference, in our case the Lennard-Jones binding energy profile from DFT, whereas $f_L$ is same binding energy profile computed with LAMMPS using the current $\epsilon$ and $\sigma$.
We used the weight function $w(r)=exp[-(\frac{r-r_0}{\zeta})^2]$ to ensure that the most relevant part, the minimum of the Lennard-Jones at $r_0=\SI{4.94}{\AA}$ and its directs surroundings, are represented correctly.
The amplitude of the relevant interval around the minimum is tuned with the $\zeta$ parameter.
Both the reference energy profile and the one from LAMMPS are obtained by fixing the interlayer distance between G and MoS$_2$, in our case we fixed the carbon atoms and the outermost sulfur atoms in the $z$ direction.
The heterostructure itself results from a 4x4 MoS$_2$ unit cell repetition and a 5x5 G unit cell repetition, in which the resulting stress of 2.6\% is applied to MoS$_2$.
The results of this optimization can be found in Tab.~\ref{tab:optim} and Fig.~\ref{fig:optim}.
\begin{table}[h!]
\begin{tabular}{ |p{2cm}||p{2cm}|p{2cm}|p{2cm}| }
 \hline
 \multicolumn{4}{|c|}{Optimized LJ Parameters} \\
 \hline
  Atoms & $\epsilon$ [meV] & $\sigma$ [$\AA$] & $\zeta$ [$\AA$] \\
 \hline
 C-S   & 1.64 & 3.640 & 0.30 \\
 C-Mo & 4.55 & 4.391 & 0.30 \\
 \hline
\end{tabular}
\caption{Optimized LJ Parameters for the interlayer interaction between G and MoS$_2$.}
\label{tab:optim}
\end{table}
\section{LJ enhancement}
\begin{figure}
  \includegraphics[width=0.89\textwidth]{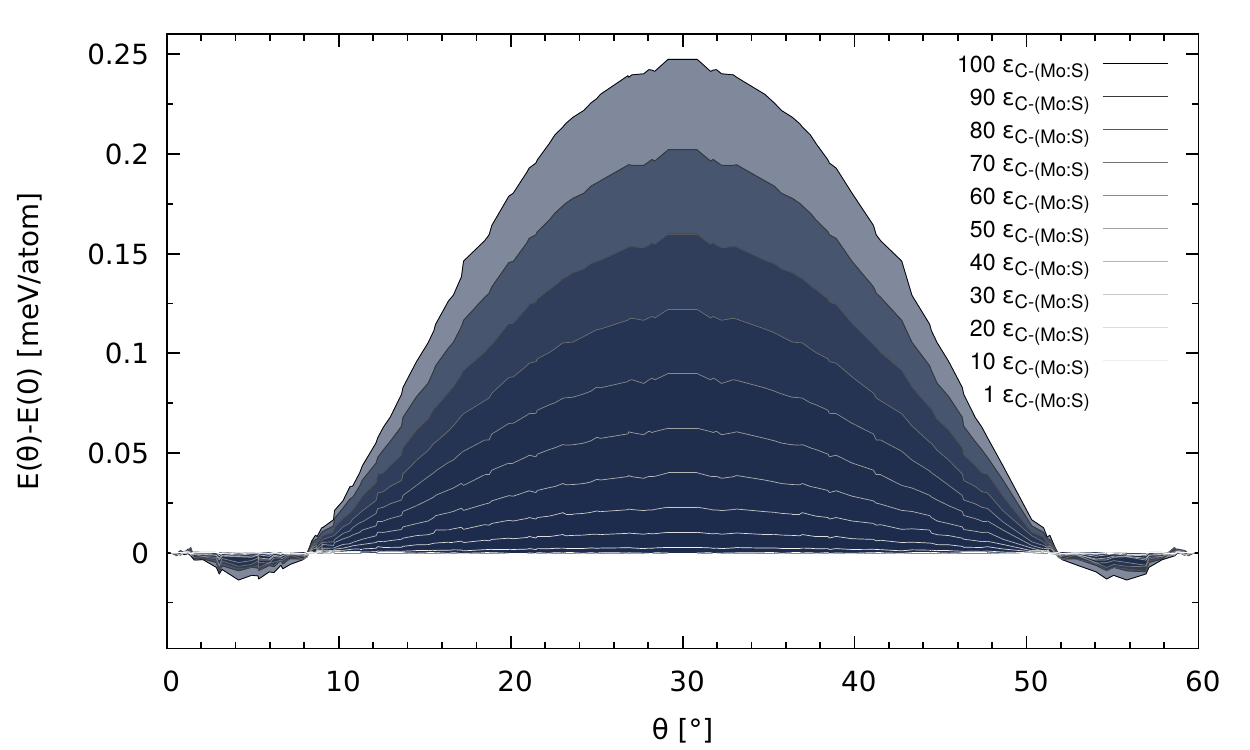}
  \caption{\label{fig:ljenhance}
   Enhancement of the LJ-coupling. Plotted are the energy in meV/atom versus angle $E(\theta)$ for rigid MoS$_2$ and soft G for increasing values of LJ coupling $f \epsilon_{C-(Mo:S}$, as reported in the legend.  
   }
\end{figure}

\begin{figure}
  \includegraphics[width=0.89\textwidth]{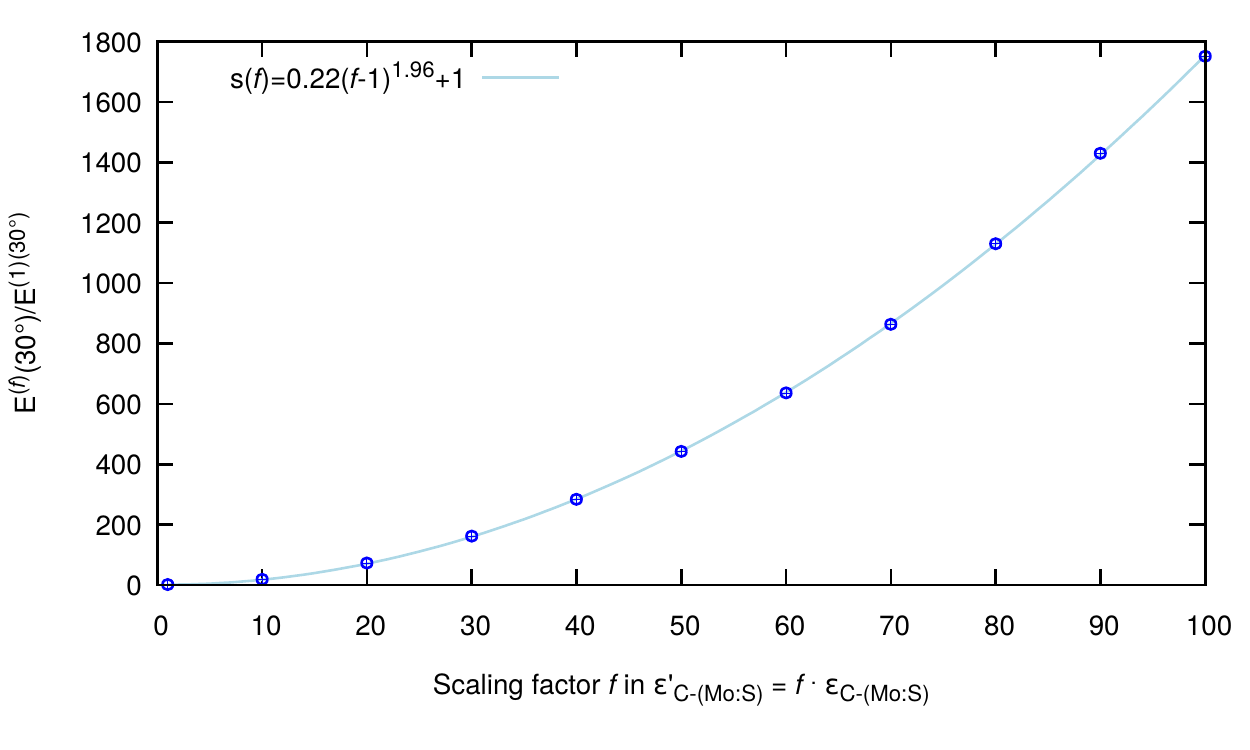}
  \caption{\label{fig:lj_scaling}
  Scaling relation between the energy and the LJ parameter $\epsilon$. Blue circles represent the energy at $\theta=\ang{30}$ computed at a given enhancement factor $f$, with respect to $f=1$, versus the scaling factor $f$. The light-blue line shows the fitted power law reported in the legend.
   }
\end{figure}

%
The LJ-coupling between the layers of MoS$_2$ and G was enhanced during the constraint simulations as mentioned in the main text.
This was done because the strain posed on the MoS$_2$ lattice, in order to create a supercell suitable for the application of PBC, results in a noise significantly affecting the energy profile upon rotating the lattices.
In Fig.~\ref{fig:ljenhance}, we report the energy profile $E(\theta)$ for different values of the scaling factor $f$ in $\epsilon'=f \epsilon$. 
It can be seen that this computational trick does not alter the physics but purely amplifies the energy trends that otherwise get progressively hidden by the noise.
Figure~\ref{fig:lj_scaling} reports the scaling relation at $\theta=\ang{30}$, showing an almost quadratic behaviour.
In order to make comparison between 
Fig. 1 and Fig. 2 in the main text 
easier, we scaled back the value computed at $\epsilon'_{LJ}=100\epsilon_{LJ}$ according to $E^{100}(\ang{30})/E^{1}(\ang{30})=1751.57$.
In other words, both the minima and maxima remain located at the same angle, however, their absolute energy value is scaled according to the LJ-coupling.
%
%
\section{Phonon dispersion and sound velocity}
\begin{figure*}
  \includegraphics[width=0.89\textwidth]{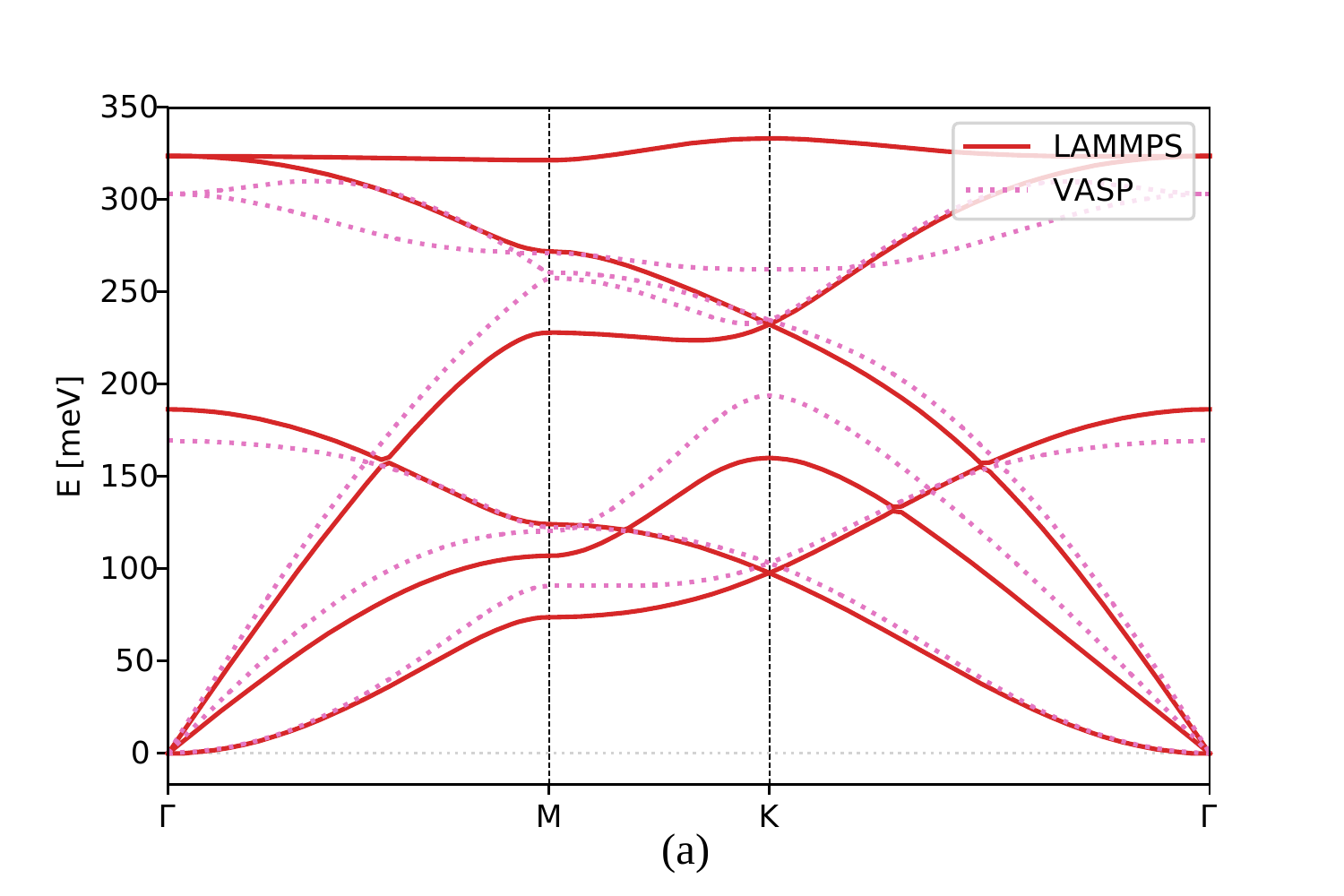}
%
  \includegraphics[width=0.89\textwidth]{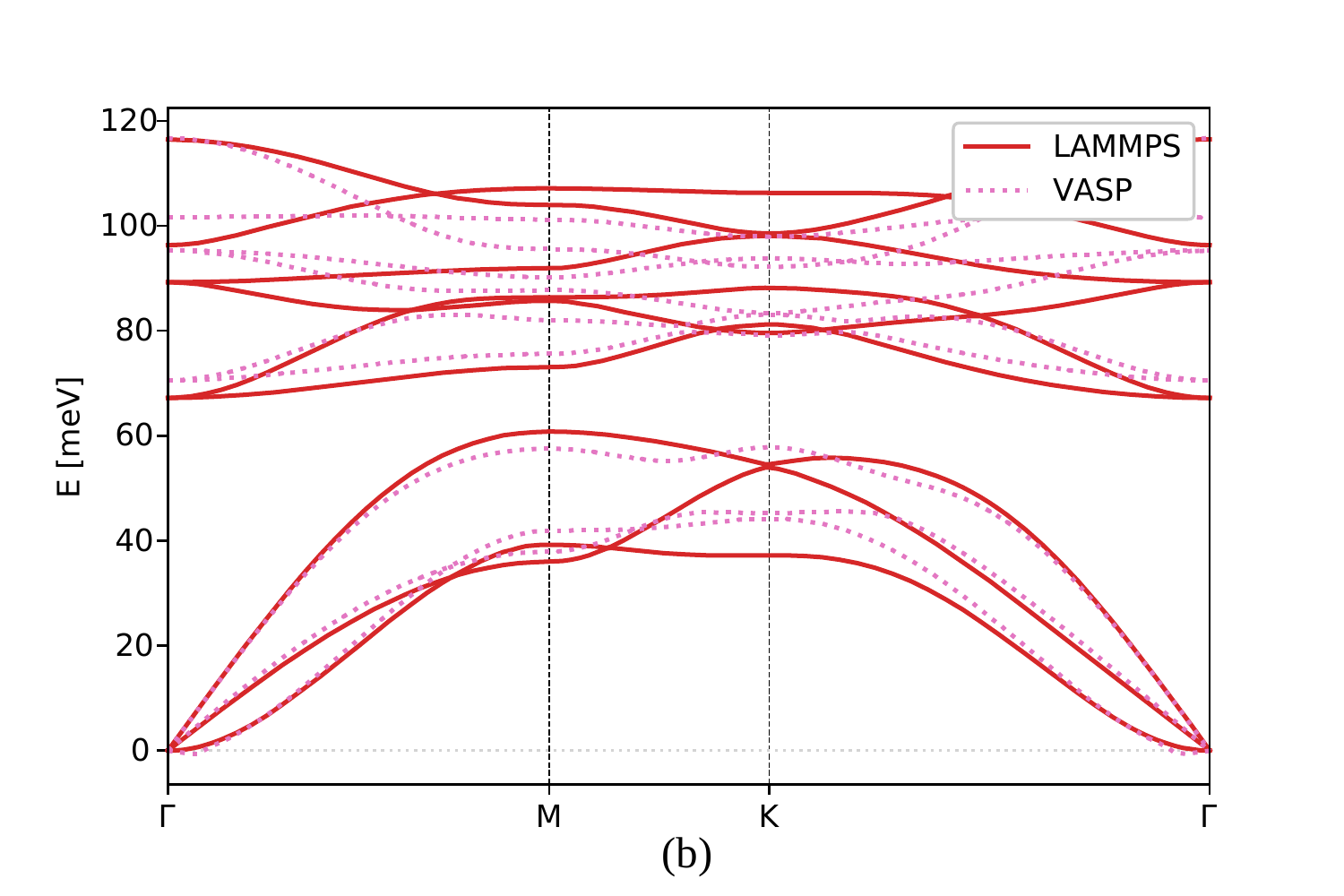}
  \caption{\label{fig:phn_comp}
  Phonon band structure of (a) G and (b) MoS$_2$ computed with LAMMPS (solid lines) and VASP (dashed lines). The $y$ axis reports the phonon energy, while the $x$ axis marks the distance from the origin along the path $\Gamma \to M \to K \to \Gamma$.}
\end{figure*}
\Cref{fig:phn_comp} reports the phonon band structure along the path $\Gamma \to M  \to K \to \Gamma$ of G and MoS$_2$, panels (a) and (b) respectively.
These figures allow us to compare phonon dispersion computed from quantum forces, at the DFT level, and from classical forces.
We focus on the acoustic modes first.
Quantum and classical dispersion are in good agreement around $\Gamma$, the center of the Brillouin zone; this means that the long wavelength distortions at the base of NM theory are well-described by the classical force fields.
As we move to the edge of the cell, towards shorter distortions, the two dispersion start to deviate.
For example, the splitting of quantum-computed transverse and longitudinal branches observed at $M$ point in G is shifted to a different $k$ in the classical results.
Similar observations can be made for the region around $K$ and for the MoS$_2$ phonon bands.
The general trend is that the classical treatment underestimates the energy of acoustic branches and overestimates the optical ones.
However, strong quantitative agreement is not needed for the qualitative statements developed in the main text and, in order to obtain the sound velocity needed as input of the NM theory, only an accurate description around the $\Gamma$ point is required.

\begin{figure}
  \includegraphics[width=0.89\textwidth]{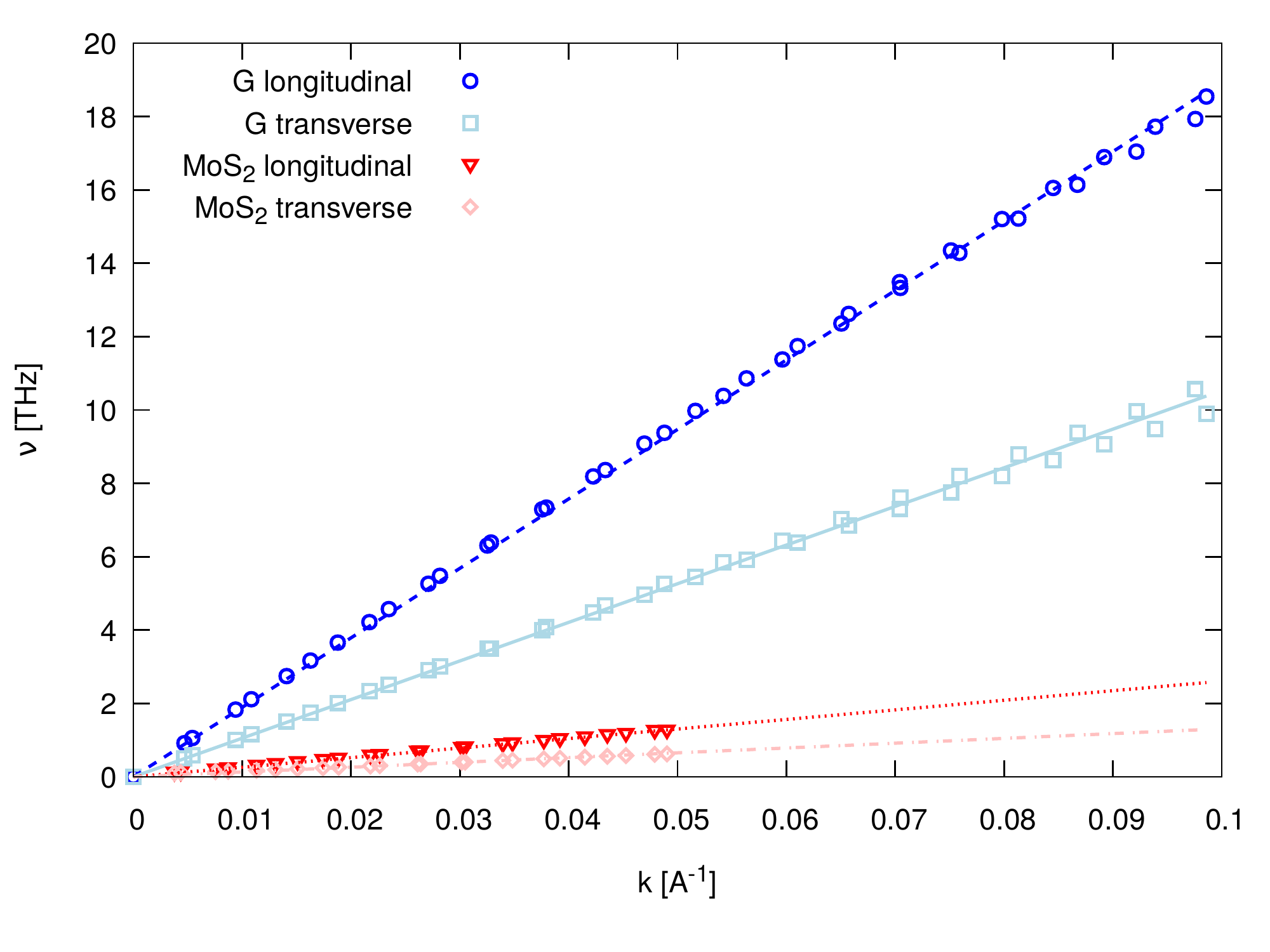}
  \caption{\label{fig:vs_combined}
  Sound velocity fit from the phonon dispersion in G (shades-of-blue symbols) and MoS$_2$ monolayer (shades-of-red symbols).
  The $y$ axis reports the frequency $\nu$ in $\SI{}{THz}$ and the $x$ axis the distance from $\Gamma$ in $\SI{}{\AA^{-1}}$, in order to make the slope $\SI{}{m/s}$.
  The color-matching lines report the linear fit of that branch, i.e. $\nu_i=v_i k$. }
\end{figure}
As explained in the main text, the  NM theory describes the distortion of a 2D layer due to interaction with rigid substrate in terms of long wavelength phonons. The theory yields analytical predictions in the limit of linear dispersion
$
    \omega_i(k)= v_i k ,
$
where $i=\rm{L},\rm{T}$ labels either the transverse (T) or longitudinal (L) branch and $v_i$ is the speed of sound of corresponding branch $i$.
This is defined as slope of the phonon dispersion near $\Gamma$: 
$
    v_i= \left. \frac{\partial \omega(k)}{\partial k}\right|_\Gamma
$.
\Cref{fig:vs_combined} shows the longitudinal and transverse branches close to $\Gamma$ of G and MoS$_2$.
The plot also shows the linear fits obtained from the points, including their fitted slopes representing the sound velocities, as reported in \Cref{tab:vs}.
\begin{table}
\begin{tabular}{l|cc}
  \hline
  Material & $v_L[\SI{}{km/s}]$ & $v_T[\SI{}{km/s}]$ \\
  \hline
  G       & 18.9403 $\pm$ 0.0005 & 10.5298 $\pm$ 0.0005 \\
  MoS$_2$ & 0.2608  $\pm$ 0.0005 & 0.131 $\pm$ 0.002 \\
  \hline
\end{tabular}
\caption{\label{tab:vs}%
  Sound velocity of transverse and longitudinal phonon branches in G and monolayer MoS$_2$ extracted from \Cref{fig:vs_combined}.
  The uncertainty arises from the linear fitting procedure.}
\end{table}
This leads to the values $\delta_{\rm{G}}=2.235$ and $\delta_{\rm{MoS_2}}=2.968$ used to evaluate eq. 4 in the main text.

\section{NM approximation limits}
As a final observation, we report the results of a heterostructure with constraints between the pure NM theory and free system.
More specifically, we consider a bilayer MoS$_2$/G where atoms are free to move in the $xy$ plane but constrained along $z$.
This corresponds to lifting the rigid substrate assumption of the NM theory, while enforcing a constant interlayer distance.
\begin{figure}
  \centering
  \includegraphics[width=0.89\textwidth]{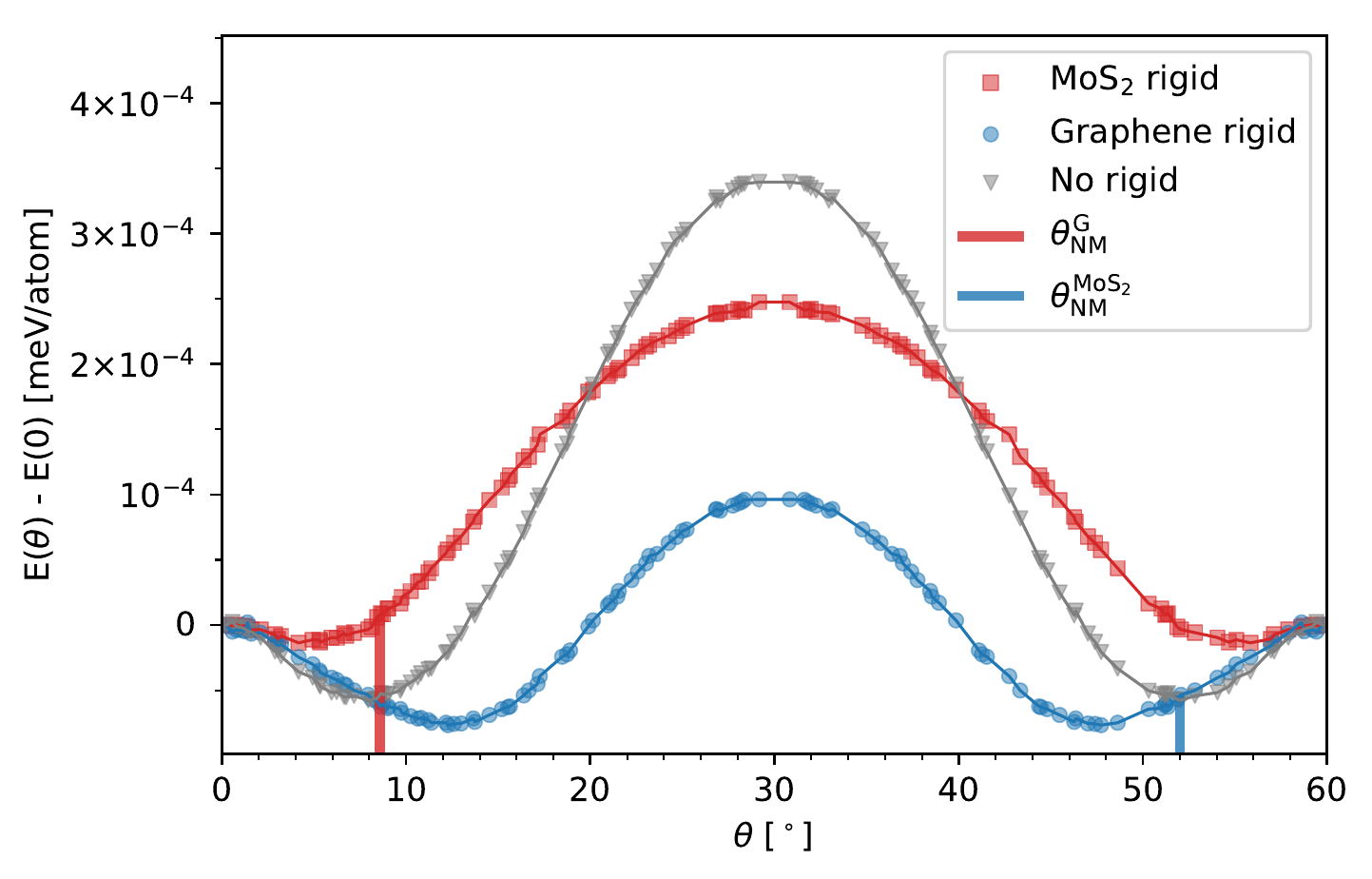}
  \caption{\label{fig:lj_novaco}
  Energy per atom $E(\theta)$, in meV/atom, as a function of the imposed angle $\theta$ in degrees for different 2D models:
  red squares refer to flexible G on top of rigid MoS$_2$; 
  blue circles refer to flexible MoS$_2$ on top of rigid G.
  Gray triangles refer to flexible MoS$_2$ on top of flexible G.
  The reference value of the energy scale is set by $E(0)$ and the values have been corrected according to the initial LJ enhancement.
  Red and blue segments mark the minimum angle predicted by the NM theory for the first and second case, respectively.
  }
  \end{figure}
Figure ~\ref{fig:lj_novaco} reports on the results of this case.
The NM theory does not cover this scenario, as here none of the two layers is rigid.
As a result, the two layers can mutually influence and distort each other, reaching configurations not included in the NM model.
A minimum at $\theta \approx \ang{8}$ is clearly visible, midway between the two rigid substrate approximations. 
This, despite the fact that the model does not describe the mutual interaction between the layers and does not provide a prediction for the the minimum-energy angle.
The behaviour of the system is thus qualitatively different from the h-BN/G heterostructures studied by Guerra et al \cite{Guerra2017}.
In that case the NM theory was found to explain quantitatively the energetics of the rigid and $z$-frozen scenario, i.e. blue and red lines in \Cref{fig:lj_novaco}, but the removing the rigid substrate constraint changed the behavior qualitatively: the minimum-energy angle predicted by NM disappeared from the energy profile of the system.
From this observation, we can conclude that relaxing the constraint of a rigid substrate in this system with a 3-dimensional single layer does not contradict the physics described by the NM model.

\nocite{*}


\begin{thebibliography}{44}%
\makeatletter
\providecommand \@ifxundefined [1]{%
 \@ifx{#1\undefined}
}%
\providecommand \@ifnum [1]{%
 \ifnum #1\expandafter \@firstoftwo
 \else \expandafter \@secondoftwo
 \fi
}%
\providecommand \@ifx [1]{%
 \ifx #1\expandafter \@firstoftwo
 \else \expandafter \@secondoftwo
 \fi
}%
\providecommand \natexlab [1]{#1}%
\providecommand \enquote  [1]{``#1''}%
\providecommand \bibnamefont  [1]{#1}%
\providecommand \bibfnamefont [1]{#1}%
\providecommand \citenamefont [1]{#1}%
\providecommand \href@noop [0]{\@secondoftwo}%
\providecommand \href [0]{\begingroup \@sanitize@url \@href}%
\providecommand \@href[1]{\@@startlink{#1}\@@href}%
\providecommand \@@href[1]{\endgroup#1\@@endlink}%
\providecommand \@sanitize@url [0]{\catcode `\\12\catcode `\$12\catcode
  `\&12\catcode `\#12\catcode `\^12\catcode `\_12\catcode `\%12\relax}%
\providecommand \@@startlink[1]{}%
\providecommand \@@endlink[0]{}%
\providecommand \url  [0]{\begingroup\@sanitize@url \@url }%
\providecommand \@url [1]{\endgroup\@href {#1}{\urlprefix }}%
\providecommand \urlprefix  [0]{URL }%
\providecommand \Eprint [0]{\href }%
\providecommand \doibase [0]{https://doi.org/}%
\providecommand \selectlanguage [0]{\@gobble}%
\providecommand \bibinfo  [0]{\@secondoftwo}%
\providecommand \bibfield  [0]{\@secondoftwo}%
\providecommand \translation [1]{[#1]}%
\providecommand \BibitemOpen [0]{}%
\providecommand \bibitemStop [0]{}%
\providecommand \bibitemNoStop [0]{.\EOS\space}%
\providecommand \EOS [0]{\spacefactor3000\relax}%
\providecommand \BibitemShut  [1]{\csname bibitem#1\endcsname}%
\let\auto@bib@innerbib\@empty
\bibitem [{\citenamefont {Mannix}\ \emph {et~al.}(2017)\citenamefont {Mannix},
  \citenamefont {Kiraly}, \citenamefont {Hersam},\ and\ \citenamefont
  {Guisinger}}]{mannix2017}%
  \BibitemOpen
  \bibfield  {author} {\bibinfo {author} {\bibfnamefont {A.~J.}\ \bibnamefont
  {Mannix}}, \bibinfo {author} {\bibfnamefont {B.}~\bibnamefont {Kiraly}},
  \bibinfo {author} {\bibfnamefont {M.~C.}\ \bibnamefont {Hersam}},\ and\
  \bibinfo {author} {\bibfnamefont {N.~P.}\ \bibnamefont {Guisinger}},\
  }\bibfield  {title} {\bibinfo {title} {{Synthesis and chemistry of elemental
  2D materials}},\ }\href {https://doi.org/10.1038/s41570-016-0014} {\bibfield
  {journal} {\bibinfo  {journal} {Nature Reviews Chemistry}\ }\textbf {\bibinfo
  {volume} {1}},\ \bibinfo {pages} {0014} (\bibinfo {year} {2017})}\BibitemShut
  {NoStop}%
\bibitem [{\citenamefont {Chhowalla}\ \emph {et~al.}(2013)\citenamefont
  {Chhowalla}, \citenamefont {Shin}, \citenamefont {Eda}, \citenamefont {Li},
  \citenamefont {Loh},\ and\ \citenamefont {Zhang}}]{Chhowalla2013}%
  \BibitemOpen
  \bibfield  {author} {\bibinfo {author} {\bibfnamefont {M.}~\bibnamefont
  {Chhowalla}}, \bibinfo {author} {\bibfnamefont {H.~S.}\ \bibnamefont {Shin}},
  \bibinfo {author} {\bibfnamefont {G.}~\bibnamefont {Eda}}, \bibinfo {author}
  {\bibfnamefont {L.-J.}\ \bibnamefont {Li}}, \bibinfo {author} {\bibfnamefont
  {K.~P.}\ \bibnamefont {Loh}},\ and\ \bibinfo {author} {\bibfnamefont
  {H.}~\bibnamefont {Zhang}},\ }\bibfield  {title} {\bibinfo {title} {{The
  chemistry of two-dimensional layered transition metal dichalcogenide
  nanosheets}},\ }\href {https://doi.org/10.1038/nchem.1589} {\bibfield
  {journal} {\bibinfo  {journal} {Nature Chemistry}\ }\textbf {\bibinfo
  {volume} {5}},\ \bibinfo {pages} {263} (\bibinfo {year} {2013})}\BibitemShut
  {NoStop}%
\bibitem [{\citenamefont {Vazirisereshk}\ \emph
  {et~al.}(2019{\natexlab{a}})\citenamefont {Vazirisereshk}, \citenamefont
  {Martini}, \citenamefont {Strubbe},\ and\ \citenamefont
  {Baykara}}]{Vazirisereshk2019}%
  \BibitemOpen
  \bibfield  {author} {\bibinfo {author} {\bibfnamefont {M.~R.}\ \bibnamefont
  {Vazirisereshk}}, \bibinfo {author} {\bibfnamefont {A.}~\bibnamefont
  {Martini}}, \bibinfo {author} {\bibfnamefont {D.~A.}\ \bibnamefont
  {Strubbe}},\ and\ \bibinfo {author} {\bibfnamefont {M.~Z.}\ \bibnamefont
  {Baykara}},\ }\bibfield  {title} {\bibinfo {title} {{Solid Lubrication with
  MoS{$_2$}: A Review}},\ }\href {https://doi.org/10.3390/lubricants7070057}
  {\bibfield  {journal} {\bibinfo  {journal} {Lubricants}\ }\textbf {\bibinfo
  {volume} {7}},\ \bibinfo {pages} {57} (\bibinfo {year}
  {2019}{\natexlab{a}})}\BibitemShut {NoStop}%
\bibitem [{\citenamefont {Radisavljevic}\ \emph {et~al.}(2011)\citenamefont
  {Radisavljevic}, \citenamefont {Radenovic}, \citenamefont {Brivio},
  \citenamefont {Giacometti},\ and\ \citenamefont {Kis}}]{radisavljevic2011}%
  \BibitemOpen
  \bibfield  {author} {\bibinfo {author} {\bibfnamefont {B.}~\bibnamefont
  {Radisavljevic}}, \bibinfo {author} {\bibfnamefont {A.}~\bibnamefont
  {Radenovic}}, \bibinfo {author} {\bibfnamefont {J.}~\bibnamefont {Brivio}},
  \bibinfo {author} {\bibfnamefont {V.}~\bibnamefont {Giacometti}},\ and\
  \bibinfo {author} {\bibfnamefont {A.}~\bibnamefont {Kis}},\ }\bibfield
  {title} {\bibinfo {title} {{Single-layer MoS{$_2$} transistors}},\ }\href
  {https://doi.org/10.1038/nnano.2010.279} {\bibfield  {journal} {\bibinfo
  {journal} {Nature Nanotechnology}\ }\textbf {\bibinfo {volume} {6}},\
  \bibinfo {pages} {147} (\bibinfo {year} {2011})}\BibitemShut {NoStop}%
\bibitem [{\citenamefont {Lopez-Sanchez}\ \emph {et~al.}(2013)\citenamefont
  {Lopez-Sanchez}, \citenamefont {Lembke}, \citenamefont {Kayci}, \citenamefont
  {Radenovic},\ and\ \citenamefont {Kis}}]{lopezsanchez2013}%
  \BibitemOpen
  \bibfield  {author} {\bibinfo {author} {\bibfnamefont {O.}~\bibnamefont
  {Lopez-Sanchez}}, \bibinfo {author} {\bibfnamefont {D.}~\bibnamefont
  {Lembke}}, \bibinfo {author} {\bibfnamefont {M.}~\bibnamefont {Kayci}},
  \bibinfo {author} {\bibfnamefont {A.}~\bibnamefont {Radenovic}},\ and\
  \bibinfo {author} {\bibfnamefont {A.}~\bibnamefont {Kis}},\ }\bibfield
  {title} {\bibinfo {title} {{Ultrasensitive photodetectors based on monolayer
  MoS{$_2$}}},\ }\href {https://doi.org/10.1038/nnano.2013.100} {\bibfield
  {journal} {\bibinfo  {journal} {Nature Nanotechnology}\ }\textbf {\bibinfo
  {volume} {8}},\ \bibinfo {pages} {497} (\bibinfo {year} {2013})}\BibitemShut
  {NoStop}%
\bibitem [{\citenamefont {Lauritsen}\ \emph {et~al.}(2004)\citenamefont
  {Lauritsen}, \citenamefont {Bollinger}, \citenamefont {L{\ae}gsgaard},
  \citenamefont {Jacobsen}, \citenamefont {N{\o}rskov}, \citenamefont
  {Clausen}, \citenamefont {Tops{\o}e},\ and\ \citenamefont
  {Besenbacher}}]{Lauritsen2004}%
  \BibitemOpen
  \bibfield  {author} {\bibinfo {author} {\bibfnamefont {J.~V.}\ \bibnamefont
  {Lauritsen}}, \bibinfo {author} {\bibfnamefont {M.~V.}\ \bibnamefont
  {Bollinger}}, \bibinfo {author} {\bibfnamefont {E.}~\bibnamefont
  {L{\ae}gsgaard}}, \bibinfo {author} {\bibfnamefont {K.~W.}\ \bibnamefont
  {Jacobsen}}, \bibinfo {author} {\bibfnamefont {J.~K.}\ \bibnamefont
  {N{\o}rskov}}, \bibinfo {author} {\bibfnamefont {B.~S.}\ \bibnamefont
  {Clausen}}, \bibinfo {author} {\bibfnamefont {H.}~\bibnamefont {Tops{\o}e}},\
  and\ \bibinfo {author} {\bibfnamefont {F.}~\bibnamefont {Besenbacher}},\
  }\bibfield  {title} {\bibinfo {title} {{Atomic-scale insight into structure
  and morphology changes of MoS{$_2$} nanoclusters in hydrotreating
  catalysts}},\ }\href {https://doi.org/10.1016/j.jcat.2003.09.015} {\bibfield
  {journal} {\bibinfo  {journal} {Journal of Catalysis}\ }\textbf {\bibinfo
  {volume} {221}},\ \bibinfo {pages} {510} (\bibinfo {year}
  {2004})}\BibitemShut {NoStop}%
\bibitem [{\citenamefont {Shi}\ \emph {et~al.}(2009{\natexlab{a}})\citenamefont
  {Shi}, \citenamefont {Jiao}, \citenamefont {Hermann},\ and\ \citenamefont
  {Wang}}]{shi2009a}%
  \BibitemOpen
  \bibfield  {author} {\bibinfo {author} {\bibfnamefont {X.-R.}\ \bibnamefont
  {Shi}}, \bibinfo {author} {\bibfnamefont {H.}~\bibnamefont {Jiao}}, \bibinfo
  {author} {\bibfnamefont {K.}~\bibnamefont {Hermann}},\ and\ \bibinfo {author}
  {\bibfnamefont {J.}~\bibnamefont {Wang}},\ }\bibfield  {title} {\bibinfo
  {title} {{CO hydrogenation reaction on sulfided molybdenum catalysts}},\
  }\href {https://doi.org/10.1016/j.molcata.2009.06.025} {\bibfield  {journal}
  {\bibinfo  {journal} {Journal of Molecular Catalysis A: Chemical}\ }\textbf
  {\bibinfo {volume} {312}},\ \bibinfo {pages} {7} (\bibinfo {year}
  {2009}{\natexlab{a}})}\BibitemShut {NoStop}%
\bibitem [{\citenamefont {Shi}\ \emph {et~al.}(2009{\natexlab{b}})\citenamefont
  {Shi}, \citenamefont {Wang}, \citenamefont {Hu}, \citenamefont {Wang},
  \citenamefont {Chen}, \citenamefont {Qin},\ and\ \citenamefont
  {Wang}}]{shi2009}%
  \BibitemOpen
  \bibfield  {author} {\bibinfo {author} {\bibfnamefont {X.-R.}\ \bibnamefont
  {Shi}}, \bibinfo {author} {\bibfnamefont {S.-G.}\ \bibnamefont {Wang}},
  \bibinfo {author} {\bibfnamefont {J.}~\bibnamefont {Hu}}, \bibinfo {author}
  {\bibfnamefont {H.}~\bibnamefont {Wang}}, \bibinfo {author} {\bibfnamefont
  {Y.-Y.}\ \bibnamefont {Chen}}, \bibinfo {author} {\bibfnamefont
  {Z.}~\bibnamefont {Qin}},\ and\ \bibinfo {author} {\bibfnamefont
  {J.}~\bibnamefont {Wang}},\ }\bibfield  {title} {\bibinfo {title} {{Density
  functional theory study on water–gas-shift reaction over molybdenum
  disulfide}},\ }\href {https://doi.org/10.1016/j.apcata.2009.05.050}
  {\bibfield  {journal} {\bibinfo  {journal} {Applied Catalysis A: General}\
  }\textbf {\bibinfo {volume} {365}},\ \bibinfo {pages} {62} (\bibinfo {year}
  {2009}{\natexlab{b}})}\BibitemShut {NoStop}%
\bibitem [{\citenamefont {Dienwiebel}\ \emph {et~al.}(2004)\citenamefont
  {Dienwiebel}, \citenamefont {Verhoeven}, \citenamefont {Pradeep},
  \citenamefont {Frenken}, \citenamefont {Heimberg},\ and\ \citenamefont
  {Zandbergen}}]{dienwiebel2004}%
  \BibitemOpen
  \bibfield  {author} {\bibinfo {author} {\bibfnamefont {M.}~\bibnamefont
  {Dienwiebel}}, \bibinfo {author} {\bibfnamefont {G.~S.}\ \bibnamefont
  {Verhoeven}}, \bibinfo {author} {\bibfnamefont {N.}~\bibnamefont {Pradeep}},
  \bibinfo {author} {\bibfnamefont {J.~W.~M.}\ \bibnamefont {Frenken}},
  \bibinfo {author} {\bibfnamefont {J.~A.}\ \bibnamefont {Heimberg}},\ and\
  \bibinfo {author} {\bibfnamefont {H.~W.}\ \bibnamefont {Zandbergen}},\
  }\bibfield  {title} {\bibinfo {title} {{Superlubricity of Graphite}},\ }\href
  {https://doi.org/10.1103/PhysRevLett.92.126101} {\bibfield  {journal}
  {\bibinfo  {journal} {Physical Review Letters}\ }\textbf {\bibinfo {volume}
  {92}},\ \bibinfo {pages} {126101} (\bibinfo {year} {2004})}\BibitemShut
  {NoStop}%
\bibitem [{\citenamefont {Vazirisereshk}\ \emph
  {et~al.}(2019{\natexlab{b}})\citenamefont {Vazirisereshk}, \citenamefont
  {Ye}, \citenamefont {Ye}, \citenamefont {Otero-de-la Roza}, \citenamefont
  {Zhao}, \citenamefont {Gao}, \citenamefont {Johnson}, \citenamefont
  {Johnson}, \citenamefont {Carpick},\ and\ \citenamefont
  {Martini}}]{Vazirisereshk2019a}%
  \BibitemOpen
  \bibfield  {author} {\bibinfo {author} {\bibfnamefont {M.~R.}\ \bibnamefont
  {Vazirisereshk}}, \bibinfo {author} {\bibfnamefont {H.}~\bibnamefont {Ye}},
  \bibinfo {author} {\bibfnamefont {Z.}~\bibnamefont {Ye}}, \bibinfo {author}
  {\bibfnamefont {A.}~\bibnamefont {Otero-de-la Roza}}, \bibinfo {author}
  {\bibfnamefont {M.-Q.}\ \bibnamefont {Zhao}}, \bibinfo {author}
  {\bibfnamefont {Z.}~\bibnamefont {Gao}}, \bibinfo {author} {\bibfnamefont
  {A.~T.~C.}\ \bibnamefont {Johnson}}, \bibinfo {author} {\bibfnamefont
  {E.~R.}\ \bibnamefont {Johnson}}, \bibinfo {author} {\bibfnamefont {R.~W.}\
  \bibnamefont {Carpick}},\ and\ \bibinfo {author} {\bibfnamefont
  {A.}~\bibnamefont {Martini}},\ }\bibfield  {title} {\bibinfo {title} {{
  Origin of Nanoscale Friction Contrast between Supported Graphene, MoS{$_2$} ,
  and a Graphene/MoS{$_2$} Heterostructure }},\ }\href
  {https://doi.org/10.1021/acs.nanolett.9b02035} {\bibfield  {journal}
  {\bibinfo  {journal} {Nano Letters}\ }\textbf {\bibinfo {volume} {19}},\
  \bibinfo {pages} {5496} (\bibinfo {year} {2019}{\natexlab{b}})}\BibitemShut
  {NoStop}%
\bibitem [{\citenamefont {Novoselov}\ \emph {et~al.}(2016)\citenamefont
  {Novoselov}, \citenamefont {Mishchenko}, \citenamefont {Carvalho},\ and\
  \citenamefont {Castro~Neto}}]{novoselov2016a}%
  \BibitemOpen
  \bibfield  {author} {\bibinfo {author} {\bibfnamefont {K.~S.}\ \bibnamefont
  {Novoselov}}, \bibinfo {author} {\bibfnamefont {A.}~\bibnamefont
  {Mishchenko}}, \bibinfo {author} {\bibfnamefont {A.}~\bibnamefont
  {Carvalho}},\ and\ \bibinfo {author} {\bibfnamefont {A.~H.}\ \bibnamefont
  {Castro~Neto}},\ }\bibfield  {title} {\bibinfo {title} {{2D materials and van
  der Waals heterostructures}},\ }\href
  {https://doi.org/10.1126/science.aac9439} {\bibfield  {journal} {\bibinfo
  {journal} {Science}\ }\textbf {\bibinfo {volume} {353}},\ \bibinfo {pages}
  {9439} (\bibinfo {year} {2016})}\BibitemShut {NoStop}%
\bibitem [{\citenamefont {Liu}\ \emph {et~al.}(2016{\natexlab{a}})\citenamefont
  {Liu}, \citenamefont {Weiss}, \citenamefont {Duan}, \citenamefont {Cheng},
  \citenamefont {Huang},\ and\ \citenamefont {Duan}}]{liu2016}%
  \BibitemOpen
  \bibfield  {author} {\bibinfo {author} {\bibfnamefont {Y.}~\bibnamefont
  {Liu}}, \bibinfo {author} {\bibfnamefont {N.~O.}\ \bibnamefont {Weiss}},
  \bibinfo {author} {\bibfnamefont {X.}~\bibnamefont {Duan}}, \bibinfo {author}
  {\bibfnamefont {H.-C.}\ \bibnamefont {Cheng}}, \bibinfo {author}
  {\bibfnamefont {Y.}~\bibnamefont {Huang}},\ and\ \bibinfo {author}
  {\bibfnamefont {X.}~\bibnamefont {Duan}},\ }\bibfield  {title} {\bibinfo
  {title} {{Van der Waals heterostructures and devices}},\ }\href
  {https://doi.org/10.1038/natrevmats.2016.42} {\bibfield  {journal} {\bibinfo
  {journal} {Nature Reviews Materials}\ }\textbf {\bibinfo {volume} {1}},\
  \bibinfo {pages} {16042} (\bibinfo {year} {2016}{\natexlab{a}})}\BibitemShut
  {NoStop}%
\bibitem [{\citenamefont {Geim}\ and\ \citenamefont
  {Grigorieva}(2013)}]{geim2013}%
  \BibitemOpen
  \bibfield  {author} {\bibinfo {author} {\bibfnamefont {A.~K.}\ \bibnamefont
  {Geim}}\ and\ \bibinfo {author} {\bibfnamefont {I.~V.}\ \bibnamefont
  {Grigorieva}},\ }\bibfield  {title} {\bibinfo {title} {{Van der Waals
  heterostructures}},\ }\href {https://doi.org/10.1038/nature12385} {\bibfield
  {journal} {\bibinfo  {journal} {Nature}\ }\textbf {\bibinfo {volume} {499}},\
  \bibinfo {pages} {419} (\bibinfo {year} {2013})}\BibitemShut {NoStop}%
  \bibitem [{\citenamefont {Li}\ \emph {et~al.}(2019)\citenamefont {Li},
  \citenamefont {Guo}, \citenamefont {Wang}, \citenamefont {Zhang},
  \citenamefont {Zhang}, \citenamefont {Chen}, \citenamefont {Fan},
  \citenamefont {Zhang}, \citenamefont {Li},\ and\ \citenamefont
  {Lau}}]{li2019}%
  \BibitemOpen
  \bibfield  {author} {\bibinfo {author} {\bibfnamefont {B.}~\bibnamefont
  {Li}}, \bibinfo {author} {\bibfnamefont {H.}~\bibnamefont {Guo}}, \bibinfo
  {author} {\bibfnamefont {Y.}~\bibnamefont {Wang}}, \bibinfo {author}
  {\bibfnamefont {W.}~\bibnamefont {Zhang}}, \bibinfo {author} {\bibfnamefont
  {Q.}~\bibnamefont {Zhang}}, \bibinfo {author} {\bibfnamefont
  {L.}~\bibnamefont {Chen}}, \bibinfo {author} {\bibfnamefont {X.}~\bibnamefont
  {Fan}}, \bibinfo {author} {\bibfnamefont {W.}~\bibnamefont {Zhang}}, \bibinfo
  {author} {\bibfnamefont {Y.}~\bibnamefont {Li}},\ and\ \bibinfo {author}
  {\bibfnamefont {W.~M.}\ \bibnamefont {Lau}},\ }\bibfield  {title} {\bibinfo
  {title} {{Asymmetric MXene/monolayer transition metal dichalcogenide
  heterostructures for functional applications}},\ }\bibfield  {journal}
  {\bibinfo  {journal} {npj Computational Materials}\ }\textbf {\bibinfo
  {volume} {5}},\ \href {https://doi.org/10.1038/s41524-019-0155-6}
  {10.1038/s41524-019-0155-6} (\bibinfo {year} {2019})\BibitemShut {NoStop}%
\bibitem [{\citenamefont {Zhu}\ \emph {et~al.}(2019)\citenamefont {Zhu},
  \citenamefont {Pochet},\ and\ \citenamefont {Johnson}}]{zhu2019}%
  \BibitemOpen
  \bibfield  {author} {\bibinfo {author} {\bibfnamefont {S.}~\bibnamefont
  {Zhu}}, \bibinfo {author} {\bibfnamefont {P.}~\bibnamefont {Pochet}},\ and\
  \bibinfo {author} {\bibfnamefont {H.~T.}\ \bibnamefont {Johnson}},\
  }\bibfield  {title} {\bibinfo {title} {{Controlling Rotation of
  Two-Dimensional Material Flakes}},\ }\href
  {https://doi.org/10.1021/acsnano.9b01794} {\bibfield  {journal} {\bibinfo
  {journal} {ACS Nano}\ }\textbf {\bibinfo {volume} {13}},\ \bibinfo {pages}
  {6925} (\bibinfo {year} {2019})}\BibitemShut {NoStop}%
\bibitem [{\citenamefont {Du}\ \emph {et~al.}(2017)\citenamefont {Du},
  \citenamefont {Yu}, \citenamefont {Liao}, \citenamefont {Wang}, \citenamefont
  {Xie}, \citenamefont {Lu}, \citenamefont {Zhu}, \citenamefont {Li},
  \citenamefont {Shen}, \citenamefont {Chen}, \citenamefont {Yang},
  \citenamefont {Shi},\ and\ \citenamefont {Zhang}}]{du2017}%
  \BibitemOpen
  \bibfield  {author} {\bibinfo {author} {\bibfnamefont {L.}~\bibnamefont
  {Du}}, \bibinfo {author} {\bibfnamefont {H.}~\bibnamefont {Yu}}, \bibinfo
  {author} {\bibfnamefont {M.}~\bibnamefont {Liao}}, \bibinfo {author}
  {\bibfnamefont {S.}~\bibnamefont {Wang}}, \bibinfo {author} {\bibfnamefont
  {L.}~\bibnamefont {Xie}}, \bibinfo {author} {\bibfnamefont {X.}~\bibnamefont
  {Lu}}, \bibinfo {author} {\bibfnamefont {J.}~\bibnamefont {Zhu}}, \bibinfo
  {author} {\bibfnamefont {N.}~\bibnamefont {Li}}, \bibinfo {author}
  {\bibfnamefont {C.}~\bibnamefont {Shen}}, \bibinfo {author} {\bibfnamefont
  {P.}~\bibnamefont {Chen}}, \bibinfo {author} {\bibfnamefont {R.}~\bibnamefont
  {Yang}}, \bibinfo {author} {\bibfnamefont {D.}~\bibnamefont {Shi}},\ and\
  \bibinfo {author} {\bibfnamefont {G.}~\bibnamefont {Zhang}},\ }\bibfield
  {title} {\bibinfo {title} {{Modulating PL and electronic structures of
  MoS{$_2$} /graphene heterostructures via interlayer twisting angle}},\ }\href
  {https://doi.org/10.1063/1.5011120} {\bibfield  {journal} {\bibinfo
  {journal} {Applied Physics Letters}\ }\textbf {\bibinfo {volume} {111}},\
  \bibinfo {pages} {263106} (\bibinfo {year} {2017})}\BibitemShut {NoStop}%
\bibitem [{\citenamefont {Huang}\ \emph {et~al.}(2015)\citenamefont {Huang},
  \citenamefont {Chen}, \citenamefont {Zhang}, \citenamefont {Quek},
  \citenamefont {Chen}, \citenamefont {Li}, \citenamefont {Hsu}, \citenamefont
  {Chang}, \citenamefont {Zheng}, \citenamefont {Chen},\ and\ \citenamefont
  {Wee}}]{lihuang2015}%
  \BibitemOpen
  \bibfield  {author} {\bibinfo {author} {\bibfnamefont {Y.~L.}\ \bibnamefont
  {Huang}}, \bibinfo {author} {\bibfnamefont {Y.}~\bibnamefont {Chen}},
  \bibinfo {author} {\bibfnamefont {W.}~\bibnamefont {Zhang}}, \bibinfo
  {author} {\bibfnamefont {S.~Y.}\ \bibnamefont {Quek}}, \bibinfo {author}
  {\bibfnamefont {C.-H.}\ \bibnamefont {Chen}}, \bibinfo {author}
  {\bibfnamefont {L.-J.}\ \bibnamefont {Li}}, \bibinfo {author} {\bibfnamefont
  {W.-T.}\ \bibnamefont {Hsu}}, \bibinfo {author} {\bibfnamefont {W.-H.}\
  \bibnamefont {Chang}}, \bibinfo {author} {\bibfnamefont {Y.~J.}\ \bibnamefont
  {Zheng}}, \bibinfo {author} {\bibfnamefont {W.}~\bibnamefont {Chen}},\ and\
  \bibinfo {author} {\bibfnamefont {A.~T.~S.}\ \bibnamefont {Wee}},\ }\bibfield
   {title} {\bibinfo {title} {{Bandgap tunability at single-layer molybdenum
  disulphide grain boundaries}},\ }\href {https://doi.org/10.1038/ncomms7298}
  {\bibfield  {journal} {\bibinfo  {journal} {Nature Communications}\ }\textbf
  {\bibinfo {volume} {6}},\ \bibinfo {pages} {6298} (\bibinfo {year}
  {2015})}\BibitemShut {NoStop}%
\bibitem [{\citenamefont {Martin}\ and\ \citenamefont
  {Erdemir}(2018)}]{martin2018}%
  \BibitemOpen
  \bibfield  {author} {\bibinfo {author} {\bibfnamefont {J.~M.}\ \bibnamefont
  {Martin}}\ and\ \bibinfo {author} {\bibfnamefont {A.}~\bibnamefont
  {Erdemir}},\ }\bibfield  {title} {\bibinfo {title} {{Superlubricity:
  Friction’s vanishing act}},\ }\href {https://doi.org/10.1063/PT.3.3897}
  {\bibfield  {journal} {\bibinfo  {journal} {Physics Today}\ }\textbf
  {\bibinfo {volume} {71}},\ \bibinfo {pages} {40} (\bibinfo {year}
  {2018})}\BibitemShut {NoStop}%
\bibitem [{\citenamefont {Cao}\ \emph {et~al.}(2018)\citenamefont {Cao},
  \citenamefont {Fatemi}, \citenamefont {Fang}, \citenamefont {Watanabe},
  \citenamefont {Taniguchi}, \citenamefont {Kaxiras},\ and\ \citenamefont
  {Jarillo-Herrero}}]{cao2018}%
  \BibitemOpen
  \bibfield  {author} {\bibinfo {author} {\bibfnamefont {Y.}~\bibnamefont
  {Cao}}, \bibinfo {author} {\bibfnamefont {V.}~\bibnamefont {Fatemi}},
  \bibinfo {author} {\bibfnamefont {S.}~\bibnamefont {Fang}}, \bibinfo {author}
  {\bibfnamefont {K.}~\bibnamefont {Watanabe}}, \bibinfo {author}
  {\bibfnamefont {T.}~\bibnamefont {Taniguchi}}, \bibinfo {author}
  {\bibfnamefont {E.}~\bibnamefont {Kaxiras}},\ and\ \bibinfo {author}
  {\bibfnamefont {P.}~\bibnamefont {Jarillo-Herrero}},\ }\bibfield  {title}
  {\bibinfo {title} {{Unconventional superconductivity in magic-angle graphene
  superlattices}},\ }\href {https://doi.org/10.1038/nature26160} {\bibfield
  {journal} {\bibinfo  {journal} {Nature}\ }\textbf {\bibinfo {volume} {556}},\
  \bibinfo {pages} {43} (\bibinfo {year} {2018})}\BibitemShut {NoStop}%
\bibitem [{\citenamefont {Liao}\ \emph {et~al.}(2018)\citenamefont {Liao},
  \citenamefont {Wu}, \citenamefont {Du}, \citenamefont {Zhang}, \citenamefont
  {Wei}, \citenamefont {Zhu}, \citenamefont {Yu}, \citenamefont {Tang},
  \citenamefont {Gu}, \citenamefont {Xing}, \citenamefont {Yang}, \citenamefont
  {Shi}, \citenamefont {Yao},\ and\ \citenamefont {Zhang}}]{liao2018}%
  \BibitemOpen
  \bibfield  {author} {\bibinfo {author} {\bibfnamefont {M.}~\bibnamefont
  {Liao}}, \bibinfo {author} {\bibfnamefont {Z.-W.}\ \bibnamefont {Wu}},
  \bibinfo {author} {\bibfnamefont {L.}~\bibnamefont {Du}}, \bibinfo {author}
  {\bibfnamefont {T.}~\bibnamefont {Zhang}}, \bibinfo {author} {\bibfnamefont
  {Z.}~\bibnamefont {Wei}}, \bibinfo {author} {\bibfnamefont {J.}~\bibnamefont
  {Zhu}}, \bibinfo {author} {\bibfnamefont {H.}~\bibnamefont {Yu}}, \bibinfo
  {author} {\bibfnamefont {J.}~\bibnamefont {Tang}}, \bibinfo {author}
  {\bibfnamefont {L.}~\bibnamefont {Gu}}, \bibinfo {author} {\bibfnamefont
  {Y.}~\bibnamefont {Xing}}, \bibinfo {author} {\bibfnamefont {R.}~\bibnamefont
  {Yang}}, \bibinfo {author} {\bibfnamefont {D.}~\bibnamefont {Shi}}, \bibinfo
  {author} {\bibfnamefont {Y.}~\bibnamefont {Yao}},\ and\ \bibinfo {author}
  {\bibfnamefont {G.}~\bibnamefont {Zhang}},\ }\bibfield  {title} {\bibinfo
  {title} {{Twist angle-dependent conductivities across MoS{$_2$}/graphene
  heterojunctions}},\ }\href {https://doi.org/10.1038/s41467-018-06555-w}
  {\bibfield  {journal} {\bibinfo  {journal} {Nature Communications}\ }\textbf
  {\bibinfo {volume} {9}},\ \bibinfo {pages} {4068} (\bibinfo {year}
  {2018})}\BibitemShut {NoStop}%
\bibitem [{\citenamefont {Liu}\ \emph {et~al.}(2016{\natexlab{b}})\citenamefont
  {Liu}, \citenamefont {Balla}, \citenamefont {Bergeron}, \citenamefont
  {Campbell}, \citenamefont {Bedzyk},\ and\ \citenamefont {Hersam}}]{Liu2016a}%
  \BibitemOpen
  \bibfield  {author} {\bibinfo {author} {\bibfnamefont {X.}~\bibnamefont
  {Liu}}, \bibinfo {author} {\bibfnamefont {I.}~\bibnamefont {Balla}}, \bibinfo
  {author} {\bibfnamefont {H.}~\bibnamefont {Bergeron}}, \bibinfo {author}
  {\bibfnamefont {G.~P.}\ \bibnamefont {Campbell}}, \bibinfo {author}
  {\bibfnamefont {M.~J.}\ \bibnamefont {Bedzyk}},\ and\ \bibinfo {author}
  {\bibfnamefont {M.~C.}\ \bibnamefont {Hersam}},\ }\bibfield  {title}
  {\bibinfo {title} {{Rotationally commensurate growth of MoS{$_2$} on
  epitaxial graphene}},\ }\href {https://doi.org/10.1021/acsnano.5b06398}
  {\bibfield  {journal} {\bibinfo  {journal} {ACS Nano}\ }\textbf {\bibinfo
  {volume} {10}},\ \bibinfo {pages} {1067} (\bibinfo {year}
  {2016}{\natexlab{b}})}\BibitemShut {NoStop}%
\bibitem [{\citenamefont {Shi}\ \emph {et~al.}(2012)\citenamefont {Shi},
  \citenamefont {Zhou}, \citenamefont {Lu}, \citenamefont {Fang}, \citenamefont
  {Lee}, \citenamefont {Hsu}, \citenamefont {Kim}, \citenamefont {Kim},
  \citenamefont {Yang}, \citenamefont {Li}, \citenamefont {Idrobo},\ and\
  \citenamefont {Kong}}]{shi2012}%
  \BibitemOpen
  \bibfield  {author} {\bibinfo {author} {\bibfnamefont {Y.}~\bibnamefont
  {Shi}}, \bibinfo {author} {\bibfnamefont {W.}~\bibnamefont {Zhou}}, \bibinfo
  {author} {\bibfnamefont {A.-Y.}\ \bibnamefont {Lu}}, \bibinfo {author}
  {\bibfnamefont {W.}~\bibnamefont {Fang}}, \bibinfo {author} {\bibfnamefont
  {Y.-H.}\ \bibnamefont {Lee}}, \bibinfo {author} {\bibfnamefont {A.~L.}\
  \bibnamefont {Hsu}}, \bibinfo {author} {\bibfnamefont {S.~M.}\ \bibnamefont
  {Kim}}, \bibinfo {author} {\bibfnamefont {K.~K.}\ \bibnamefont {Kim}},
  \bibinfo {author} {\bibfnamefont {H.~Y.}\ \bibnamefont {Yang}}, \bibinfo
  {author} {\bibfnamefont {L.-J.}\ \bibnamefont {Li}}, \bibinfo {author}
  {\bibfnamefont {J.-C.}\ \bibnamefont {Idrobo}},\ and\ \bibinfo {author}
  {\bibfnamefont {J.}~\bibnamefont {Kong}},\ }\bibfield  {title} {\bibinfo
  {title} {{van der Waals Epitaxy of MoS{$_2$} Layers Using Graphene As Growth
  Templates}},\ }\href {https://doi.org/10.1021/nl204562j} {\bibfield
  {journal} {\bibinfo  {journal} {Nano Letters}\ }\textbf {\bibinfo {volume}
  {12}},\ \bibinfo {pages} {2784} (\bibinfo {year} {2012})}\BibitemShut
  {NoStop}%
\bibitem [{\citenamefont {Lu}\ \emph {et~al.}(2015)\citenamefont {Lu},
  \citenamefont {Butler}, \citenamefont {Huang}, \citenamefont {Hsing},
  \citenamefont {Yang}, \citenamefont {Chu}, \citenamefont {Luo}, \citenamefont
  {Sun}, \citenamefont {Hsu}, \citenamefont {Yang}, \citenamefont {Wei},
  \citenamefont {Li},\ and\ \citenamefont {Lin}}]{lu2015}%
  \BibitemOpen
  \bibfield  {author} {\bibinfo {author} {\bibfnamefont {C.~I.}\ \bibnamefont
  {Lu}}, \bibinfo {author} {\bibfnamefont {C.~J.}\ \bibnamefont {Butler}},
  \bibinfo {author} {\bibfnamefont {J.~K.}\ \bibnamefont {Huang}}, \bibinfo
  {author} {\bibfnamefont {C.~R.}\ \bibnamefont {Hsing}}, \bibinfo {author}
  {\bibfnamefont {H.~H.}\ \bibnamefont {Yang}}, \bibinfo {author}
  {\bibfnamefont {Y.~H.}\ \bibnamefont {Chu}}, \bibinfo {author} {\bibfnamefont
  {C.~H.}\ \bibnamefont {Luo}}, \bibinfo {author} {\bibfnamefont {Y.~C.}\
  \bibnamefont {Sun}}, \bibinfo {author} {\bibfnamefont {S.~H.}\ \bibnamefont
  {Hsu}}, \bibinfo {author} {\bibfnamefont {K.~H.~O.}\ \bibnamefont {Yang}},
  \bibinfo {author} {\bibfnamefont {C.~M.}\ \bibnamefont {Wei}}, \bibinfo
  {author} {\bibfnamefont {L.~J.}\ \bibnamefont {Li}},\ and\ \bibinfo {author}
  {\bibfnamefont {M.~T.}\ \bibnamefont {Lin}},\ }\bibfield  {title} {\bibinfo
  {title} {{Graphite edge controlled registration of monolayer MoS{$_2$}
  crystal orientation}},\ }\href {https://doi.org/10.1063/1.4919923} {\bibfield
   {journal} {\bibinfo  {journal} {Applied Physics Letters}\ }\textbf {\bibinfo
  {volume} {106}},\ \bibinfo {pages} {2} (\bibinfo {year} {2015})}\BibitemShut
  {NoStop}%
\bibitem [{\citenamefont {Adrian}\ \emph {et~al.}(2016)\citenamefont {Adrian},
  \citenamefont {Senftleben}, \citenamefont {Morgenstern},\ and\ \citenamefont
  {Baumert}}]{adrian2016}%
  \BibitemOpen
  \bibfield  {author} {\bibinfo {author} {\bibfnamefont {M.}~\bibnamefont
  {Adrian}}, \bibinfo {author} {\bibfnamefont {A.}~\bibnamefont {Senftleben}},
  \bibinfo {author} {\bibfnamefont {S.}~\bibnamefont {Morgenstern}},\ and\
  \bibinfo {author} {\bibfnamefont {T.}~\bibnamefont {Baumert}},\ }\bibfield
  {title} {\bibinfo {title} {{Complete analysis of a transmission electron
  diffraction pattern of a MoS{$_2$}-graphite heterostructure}},\ }\href
  {https://doi.org/10.1016/j.ultramic.2016.04.002} {\bibfield  {journal}
  {\bibinfo  {journal} {Ultramicroscopy}\ }\textbf {\bibinfo {volume} {166}},\
  \bibinfo {pages} {9} (\bibinfo {year} {2016})}\BibitemShut {NoStop}%
\bibitem [{\citenamefont {Wang}\ \emph {et~al.}(2017)\citenamefont {Wang},
  \citenamefont {Xiao}, \citenamefont {Zhu}, \citenamefont {Li}, \citenamefont
  {Alsaid}, \citenamefont {Fong}, \citenamefont {Zhou}, \citenamefont {Wang},
  \citenamefont {Shi}, \citenamefont {Wang}, \citenamefont {Zettl},
  \citenamefont {Reed},\ and\ \citenamefont {Zhang}}]{Wang2017}%
  \BibitemOpen
  \bibfield  {author} {\bibinfo {author} {\bibfnamefont {Y.}~\bibnamefont
  {Wang}}, \bibinfo {author} {\bibfnamefont {J.}~\bibnamefont {Xiao}}, \bibinfo
  {author} {\bibfnamefont {H.}~\bibnamefont {Zhu}}, \bibinfo {author}
  {\bibfnamefont {Y.}~\bibnamefont {Li}}, \bibinfo {author} {\bibfnamefont
  {Y.}~\bibnamefont {Alsaid}}, \bibinfo {author} {\bibfnamefont {K.~Y.}\
  \bibnamefont {Fong}}, \bibinfo {author} {\bibfnamefont {Y.}~\bibnamefont
  {Zhou}}, \bibinfo {author} {\bibfnamefont {S.}~\bibnamefont {Wang}}, \bibinfo
  {author} {\bibfnamefont {W.}~\bibnamefont {Shi}}, \bibinfo {author}
  {\bibfnamefont {Y.}~\bibnamefont {Wang}}, \bibinfo {author} {\bibfnamefont
  {A.}~\bibnamefont {Zettl}}, \bibinfo {author} {\bibfnamefont {E.~J.}\
  \bibnamefont {Reed}},\ and\ \bibinfo {author} {\bibfnamefont
  {X.}~\bibnamefont {Zhang}},\ }\bibfield  {title} {\bibinfo {title}
  {{Structural phase transition in monolayer MoTe{$_2$} driven by electrostatic
  doping}},\ }\href {https://doi.org/10.1038/nature24043} {\bibfield  {journal}
  {\bibinfo  {journal} {Nature}\ }\textbf {\bibinfo {volume} {550}},\ \bibinfo
  {pages} {487} (\bibinfo {year} {2017})}\BibitemShut {NoStop}%
\bibitem [{\citenamefont {Wang}\ \emph {et~al.}(2015)\citenamefont {Wang},
  \citenamefont {Ma}, \citenamefont {Hu}, \citenamefont {Xu},\ and\
  \citenamefont {Wang}}]{Wang2015}%
  \BibitemOpen
  \bibfield  {author} {\bibinfo {author} {\bibfnamefont {Z.~J.}\ \bibnamefont
  {Wang}}, \bibinfo {author} {\bibfnamefont {T.~B.}\ \bibnamefont {Ma}},
  \bibinfo {author} {\bibfnamefont {Y.~Z.}\ \bibnamefont {Hu}}, \bibinfo
  {author} {\bibfnamefont {L.}~\bibnamefont {Xu}},\ and\ \bibinfo {author}
  {\bibfnamefont {H.}~\bibnamefont {Wang}},\ }\bibfield  {title} {\bibinfo
  {title} {{Energy dissipation of atomic-scale friction based on
  one-dimensional Prandtl-Tomlinson model}},\ }\href
  {https://doi.org/10.1007/s40544-015-0086-2} {\bibfield  {journal} {\bibinfo
  {journal} {Friction}\ }\textbf {\bibinfo {volume} {3}},\ \bibinfo {pages}
  {170} (\bibinfo {year} {2015})}\BibitemShut {NoStop}%
\bibitem [{\citenamefont {Ding}\ \emph {et~al.}(2016)\citenamefont {Ding},
  \citenamefont {Pei}, \citenamefont {Jiang}, \citenamefont {Huang},\ and\
  \citenamefont {Zhang}}]{Ding2016}%
  \BibitemOpen
  \bibfield  {author} {\bibinfo {author} {\bibfnamefont {Z.}~\bibnamefont
  {Ding}}, \bibinfo {author} {\bibfnamefont {Q.~X.}\ \bibnamefont {Pei}},
  \bibinfo {author} {\bibfnamefont {J.~W.}\ \bibnamefont {Jiang}}, \bibinfo
  {author} {\bibfnamefont {W.}~\bibnamefont {Huang}},\ and\ \bibinfo {author}
  {\bibfnamefont {Y.~W.}\ \bibnamefont {Zhang}},\ }\bibfield  {title} {\bibinfo
  {title} {{Interfacial thermal conductance in graphene/MoS{{$_2$}}
  heterostructures}},\ }\href {https://doi.org/10.1016/j.carbon.2015.10.046}
  {\bibfield  {journal} {\bibinfo  {journal} {Carbon}\ }\textbf {\bibinfo
  {volume} {96}},\ \bibinfo {pages} {888} (\bibinfo {year} {2016})}\BibitemShut
  {NoStop}%
\bibitem [{\citenamefont {Brenner}\ \emph {et~al.}(2002)\citenamefont
  {Brenner}, \citenamefont {{et al.}}, \citenamefont {Shenderova},
  \citenamefont {Harrison}, \citenamefont {Stuart}, \citenamefont {Ni},\ and\
  \citenamefont {Sinnott}}]{Brenner2002}%
  \BibitemOpen
  \bibfield  {author} {\bibinfo {author} {\bibfnamefont {D.~W.}\ \bibnamefont
  {Brenner}}, \bibinfo {author} {\bibnamefont {{et al.}}}, \bibinfo {author}
  {\bibfnamefont {O.~A.}\ \bibnamefont {Shenderova}}, \bibinfo {author}
  {\bibfnamefont {J.~A.}\ \bibnamefont {Harrison}}, \bibinfo {author}
  {\bibfnamefont {S.~J.}\ \bibnamefont {Stuart}}, \bibinfo {author}
  {\bibfnamefont {B.}~\bibnamefont {Ni}},\ and\ \bibinfo {author}
  {\bibfnamefont {S.~B.}\ \bibnamefont {Sinnott}},\ }\bibfield  {title}
  {\bibinfo {title} {{A second-generation reactive empirical bond order (REBO)
  potential energy expression for hydrocarbons}},\ }\href
  {http://stacks.iop.org/0953-8984/14/i=4/a=312} {\bibfield  {journal}
  {\bibinfo  {journal} {Journal of Physics-Condensed Matter}\ }\textbf
  {\bibinfo {volume} {14}},\ \bibinfo {pages} {783} (\bibinfo {year}
  {2002})}\BibitemShut {NoStop}%
\bibitem [{\citenamefont {Kresse}\ and\ \citenamefont
  {Hafner}(1993)}]{Kresse1993}%
  \BibitemOpen
  \bibfield  {author} {\bibinfo {author} {\bibfnamefont {G.}~\bibnamefont
  {Kresse}}\ and\ \bibinfo {author} {\bibfnamefont {J.}~\bibnamefont
  {Hafner}},\ }\bibfield  {title} {\bibinfo {title} {{Ab initio molecular
  dynamics for open-shell transition metals}},\ }\href
  {https://doi.org/10.1103/PhysRevB.48.13115} {\bibfield  {journal} {\bibinfo
  {journal} {Physical Review B}\ }\textbf {\bibinfo {volume} {48}},\ \bibinfo
  {pages} {13115} (\bibinfo {year} {1993})}\BibitemShut {NoStop}%
\bibitem [{\citenamefont {Kresse}\ and\ \citenamefont
  {Joubert}(1999)}]{Kresse1999}%
  \BibitemOpen
  \bibfield  {author} {\bibinfo {author} {\bibfnamefont {G.}~\bibnamefont
  {Kresse}}\ and\ \bibinfo {author} {\bibfnamefont {D.}~\bibnamefont
  {Joubert}},\ }\bibfield  {title} {\bibinfo {title} {{From ultrasoft
  pseudopotentials to the projector augmented-wave method}},\ }\href
  {https://doi.org/10.1103/PhysRevB.59.1758} {\bibfield  {journal} {\bibinfo
  {journal} {Physical Review B}\ }\textbf {\bibinfo {volume} {59}},\ \bibinfo
  {pages} {1758} (\bibinfo {year} {1999})}\BibitemShut {NoStop}%
\bibitem [{\citenamefont {Bl{\"{o}}chl}(1994)}]{Blochl1994}%
  \BibitemOpen
  \bibfield  {author} {\bibinfo {author} {\bibfnamefont {P.~E.}\ \bibnamefont
  {Bl{\"{o}}chl}},\ }\bibfield  {title} {\bibinfo {title} {{Projector
  augmented-wave method}},\ }\href {https://doi.org/10.1103/PhysRevB.50.17953}
  {\bibfield  {journal} {\bibinfo  {journal} {Physical Review B}\ }\textbf
  {\bibinfo {volume} {50}},\ \bibinfo {pages} {17953} (\bibinfo {year}
  {1994})}\BibitemShut {NoStop}%
\bibitem [{\citenamefont {Perdew}\ \emph {et~al.}(1996)\citenamefont {Perdew},
  \citenamefont {Burke},\ and\ \citenamefont {Ernzerhof}}]{Perdew1996}%
  \BibitemOpen
  \bibfield  {author} {\bibinfo {author} {\bibfnamefont {J.~P.}\ \bibnamefont
  {Perdew}}, \bibinfo {author} {\bibfnamefont {K.}~\bibnamefont {Burke}},\ and\
  \bibinfo {author} {\bibfnamefont {M.}~\bibnamefont {Ernzerhof}},\ }\bibfield
  {title} {\bibinfo {title} {{Generalized Gradient Approximation Made
  Simple}},\ }\href {https://doi.org/10.1103/PhysRevLett.77.3865} {\bibfield
  {journal} {\bibinfo  {journal} {Physical Review Letters}\ }\textbf {\bibinfo
  {volume} {77}},\ \bibinfo {pages} {3865} (\bibinfo {year}
  {1996})}\BibitemShut {NoStop}%
\bibitem [{\citenamefont {Grimme}(2006)}]{Grimme2006}%
  \BibitemOpen
  \bibfield  {author} {\bibinfo {author} {\bibfnamefont {S.}~\bibnamefont
  {Grimme}},\ }\bibfield  {title} {\bibinfo {title} {{Semiempirical GGA-type
  density functional constructed with a long-range dispersion correction}},\
  }\href {https://doi.org/10.1002/jcc.20495} {\bibfield  {journal} {\bibinfo
  {journal} {Journal of Computational Chemistry}\ }\textbf {\bibinfo {volume}
  {27}},\ \bibinfo {pages} {1787} (\bibinfo {year} {2006})}\BibitemShut
  {NoStop}%
\bibitem [{\citenamefont {Plimpton}(1995)}]{Plimpton1995}%
  \BibitemOpen
  \bibfield  {author} {\bibinfo {author} {\bibfnamefont {S.}~\bibnamefont
  {Plimpton}},\ }\bibfield  {title} {\bibinfo {title} {{Fast parallel
  algorithms for short-range molecular dynamics}},\ }\href
  {https://doi.org/10.1006/jcph.1995.1039} {\bibfield  {journal} {\bibinfo
  {journal} {Journal of Computational Physics}\ }\textbf {\bibinfo {volume}
  {117}},\ \bibinfo {pages} {1} (\bibinfo {year} {1995})}\BibitemShut {NoStop}%
\bibitem [{\citenamefont {Novaco}\ and\ \citenamefont
  {McTague}(1977)}]{novaco1977}%
  \BibitemOpen
  \bibfield  {author} {\bibinfo {author} {\bibfnamefont {A.~D.}\ \bibnamefont
  {Novaco}}\ and\ \bibinfo {author} {\bibfnamefont {J.~P.}\ \bibnamefont
  {McTague}},\ }\bibfield  {title} {\bibinfo {title} {{Orientational
  epitaxy-the orientational ordering of incommensurate structures}},\ }\href
  {https://doi.org/10.1103/PhysRevLett.38.1286} {\bibfield  {journal} {\bibinfo
   {journal} {Physical Review Letters}\ }\textbf {\bibinfo {volume} {38}},\
  \bibinfo {pages} {1286} (\bibinfo {year} {1977})}\BibitemShut {NoStop}%
\bibitem [{\citenamefont {McTague}\ and\ \citenamefont
  {Novaco}(1979)}]{mctague1979}%
  \BibitemOpen
  \bibfield  {author} {\bibinfo {author} {\bibfnamefont {J.~P.}\ \bibnamefont
  {McTague}}\ and\ \bibinfo {author} {\bibfnamefont {A.~D.}\ \bibnamefont
  {Novaco}},\ }\bibfield  {title} {\bibinfo {title} {{Substrate-induced strain
  and orientational ordering in adsorbed monolayers}},\ }\href
  {https://doi.org/10.1103/PhysRevB.19.5299} {\bibfield  {journal} {\bibinfo
  {journal} {Physical Review B}\ }\textbf {\bibinfo {volume} {19}},\ \bibinfo
  {pages} {5299} (\bibinfo {year} {1979})}\BibitemShut {NoStop}%
\bibitem [{\citenamefont {Brazda}\ \emph {et~al.}(2018)\citenamefont {Brazda},
  \citenamefont {Silva}, \citenamefont {Manini}, \citenamefont {Vanossi},
  \citenamefont {Guerra}, \citenamefont {Tosatti},\ and\ \citenamefont
  {Bechinger}}]{Brazda2018a}%
  \BibitemOpen
  \bibfield  {author} {\bibinfo {author} {\bibfnamefont {T.}~\bibnamefont
  {Brazda}}, \bibinfo {author} {\bibfnamefont {A.}~\bibnamefont {Silva}},
  \bibinfo {author} {\bibfnamefont {N.}~\bibnamefont {Manini}}, \bibinfo
  {author} {\bibfnamefont {A.}~\bibnamefont {Vanossi}}, \bibinfo {author}
  {\bibfnamefont {R.}~\bibnamefont {Guerra}}, \bibinfo {author} {\bibfnamefont
  {E.}~\bibnamefont {Tosatti}},\ and\ \bibinfo {author} {\bibfnamefont
  {C.}~\bibnamefont {Bechinger}},\ }\bibfield  {title} {\bibinfo {title}
  {{Experimental Observation of the Aubry Transition in Two-Dimensional
  Colloidal Monolayers}},\ }\href {https://doi.org/10.1103/PhysRevX.8.011050}
  {\bibfield  {journal} {\bibinfo  {journal} {Physical Review X}\ }\textbf
  {\bibinfo {volume} {8}},\ \bibinfo {pages} {011050} (\bibinfo {year}
  {2018})}\BibitemShut {NoStop}%
\bibitem [{\citenamefont {Panizon}\ \emph {et~al.}(2017)\citenamefont
  {Panizon}, \citenamefont {Guerra},\ and\ \citenamefont
  {Tosatti}}]{Panizon2017a}%
  \BibitemOpen
  \bibfield  {author} {\bibinfo {author} {\bibfnamefont {E.}~\bibnamefont
  {Panizon}}, \bibinfo {author} {\bibfnamefont {R.}~\bibnamefont {Guerra}},\
  and\ \bibinfo {author} {\bibfnamefont {E.}~\bibnamefont {Tosatti}},\
  }\bibfield  {title} {\bibinfo {title} {{Ballistic thermophoresis of
  adsorbates on free-standing graphene}},\ }\href
  {https://doi.org/10.1073/pnas.1708098114} {\bibfield  {journal} {\bibinfo
  {journal} {Proceedings of the National Academy of Sciences}\ }\textbf
  {\bibinfo {volume} {114}},\ \bibinfo {pages} {E7035} (\bibinfo {year}
  {2017})}\BibitemShut {NoStop}%
\bibitem [{\citenamefont {Guerra}\ \emph {et~al.}(2017)\citenamefont {Guerra},
  \citenamefont {van Wijk}, \citenamefont {Vanossi}, \citenamefont {Fasolino},\
  and\ \citenamefont {Tosatti}}]{Guerra2017}%
  \BibitemOpen
  \bibfield  {author} {\bibinfo {author} {\bibfnamefont {R.}~\bibnamefont
  {Guerra}}, \bibinfo {author} {\bibfnamefont {M.}~\bibnamefont {van Wijk}},
  \bibinfo {author} {\bibfnamefont {A.}~\bibnamefont {Vanossi}}, \bibinfo
  {author} {\bibfnamefont {A.}~\bibnamefont {Fasolino}},\ and\ \bibinfo
  {author} {\bibfnamefont {E.}~\bibnamefont {Tosatti}},\ }\bibfield  {title}
  {\bibinfo {title} {{Graphene on h-BN: to align or not to align?}},\ }\href
  {https://doi.org/10.1039/C7NR02352A} {\bibfield  {journal} {\bibinfo
  {journal} {Nanoscale}\ }\textbf {\bibinfo {volume} {9}},\ \bibinfo {pages}
  {8799} (\bibinfo {year} {2017})}\BibitemShut {NoStop}%
\bibitem [{\citenamefont {Mandelli}\ \emph {et~al.}(2015)\citenamefont
  {Mandelli}, \citenamefont {Vanossi}, \citenamefont {Manini},\ and\
  \citenamefont {Tosatti}}]{mandelli2015}%
  \BibitemOpen
  \bibfield  {author} {\bibinfo {author} {\bibfnamefont {D.}~\bibnamefont
  {Mandelli}}, \bibinfo {author} {\bibfnamefont {A.}~\bibnamefont {Vanossi}},
  \bibinfo {author} {\bibfnamefont {N.}~\bibnamefont {Manini}},\ and\ \bibinfo
  {author} {\bibfnamefont {E.}~\bibnamefont {Tosatti}},\ }\bibfield  {title}
  {\bibinfo {title} {{Friction Boosted by Equilibrium Misalignment of
  Incommensurate Two-Dimensional Colloid Monolayers}},\ }\href
  {https://doi.org/10.1103/PhysRevLett.114.108302} {\bibfield  {journal}
  {\bibinfo  {journal} {Physical Review Letters}\ }\textbf {\bibinfo {volume}
  {114}},\ \bibinfo {pages} {108302} (\bibinfo {year} {2015})}\BibitemShut
  {NoStop}%
\bibitem [{\citenamefont {Nelder}\ and\ \citenamefont
  {Mead}(1965)}]{Nelder1965}%
  \BibitemOpen
  \bibfield  {author} {\bibinfo {author} {\bibfnamefont {J.~A.}\ \bibnamefont
  {Nelder}}\ and\ \bibinfo {author} {\bibfnamefont {R.}~\bibnamefont {Mead}},\
  }\bibfield  {title} {\bibinfo {title} {{A Simplex Method for Function
  Minimization}},\ }\href {https://doi.org/10.1093/comjnl/7.4.308} {\bibfield
  {journal} {\bibinfo  {journal} {The Computer Journal}\ }\textbf {\bibinfo
  {volume} {7}},\ \bibinfo {pages} {308} (\bibinfo {year} {1965})}\BibitemShut
  {NoStop}%
\bibitem [{\citenamefont {Nicolini}\ and\ \citenamefont
  {Polcar}(2016)}]{Nicolini2016}%
  \BibitemOpen
  \bibfield  {author} {\bibinfo {author} {\bibfnamefont {P.}~\bibnamefont
  {Nicolini}}\ and\ \bibinfo {author} {\bibfnamefont {T.}~\bibnamefont
  {Polcar}},\ }\bibfield  {title} {\bibinfo {title} {{A comparison of empirical
  potentials for sliding simulations of MoS{$_2$}}},\ }\href
  {https://doi.org/10.1016/j.commatsci.2016.01.013} {\bibfield  {journal}
  {\bibinfo  {journal} {Computational Materials Science}\ }\textbf {\bibinfo
  {volume} {115}},\ \bibinfo {pages} {158} (\bibinfo {year}
  {2016})}\BibitemShut {NoStop}%
\bibitem [{\citenamefont {Oliphant}(2007)}]{Oliphant2007scipy}%
  \BibitemOpen
  \bibfield  {author} {\bibinfo {author} {\bibfnamefont {T.~E.}\ \bibnamefont
  {Oliphant}},\ }\href {https://doi.org/10.1109/MCSE.2007.58} {\bibinfo {title}
  {{SciPy: Open source scientific tools for Python}}} (\bibinfo {year}
  {2007})\BibitemShut {NoStop}%
\bibitem [{\citenamefont {Togo}\ and\ \citenamefont {Tanaka}(2015)}]{Togo2015}%
  \BibitemOpen
  \bibfield  {author} {\bibinfo {author} {\bibfnamefont {A.}~\bibnamefont
  {Togo}}\ and\ \bibinfo {author} {\bibfnamefont {I.}~\bibnamefont {Tanaka}},\
  }\bibfield  {title} {\bibinfo {title} {{First principles phonon calculations
  in materials science}},\ }\href
  {https://doi.org/10.1016/j.scriptamat.2015.07.021} {\bibfield  {journal}
  {\bibinfo  {journal} {Scripta Materialia}\ }\textbf {\bibinfo {volume}
  {108}},\ \bibinfo {pages} {1} (\bibinfo {year} {2015})}\BibitemShut {NoStop}%
\bibitem [{\citenamefont {Carreras}(2019)}]{carreras2019}%
  \BibitemOpen
  \bibfield  {author} {\bibinfo {author} {\bibfnamefont {A.}~\bibnamefont
  {Carreras}},\ }\href {https://github.com/abelcarreras/phonolammps} {\bibinfo
  {title} {{phonoLAMMPS}}} (\bibinfo {year} {2019})\BibitemShut {NoStop}%
\end{thebibliography}
\end{document}